# Ultra-high-density double-atom catalyst with spin moment as activity descriptor for oxygen reduction reaction

Peng Lv[1], Wenjing Lv[1], Donghai Wu[1], Gang Tang[2], Xunwang Yan[3], Zhansheng Lu[4*], Dongwei Ma[1*]

[1] *Key Laboratory for Special Functional Materials of Ministry of Education, and School of Materials Science and Engineering, Henan University, Kaifeng 475004, China*

[2] *Advanced Research Institute of Multidisciplinary Science, Beijing Institute of Technology, Beijing 100081, China*

[3] *College of Physics and Engineering, Qufu Normal University, Qufu 273165, China*

[4] *School of Physics, Henan Normal University, Xinxiang 453007, China*

**ABSTRACT**

One of the great challenges facing atomically dispersed catalysts, including single-atom catalyst (SAC) and double-atom catalyst (DAC) is their ultra-low metal loading (typically less than 5 wt%), basically limiting the practical catalytic application, such as oxygen reduction reaction (ORR) crucial to hydrogen fuel cell and metal-air battery. Although some important progresses have been achieved on ultra-high-density (UHD) SACs, the reports on UHD-DACs with stable uniform dispersion is still lacking. Herein, based on the experimentally synthesized $M_2N_6$ motif (M = Sc-Zn), we theoretically demonstrated the existence of the UHD-DACs with the metal loading > 40 wt%, which were confirmed by systematic analysis of dynamic, thermal, mechanical, thermodynamic, and electrochemical stabilities. Furthermore, ORR activities of the UHD-DACs are comparable with or even better than those of the experimentally synthesized low-density (LD) counterparts, and the $Fe_2N_6$ and $Co_2N_6$ UHD-DACs locate at the peak of the activity volcano with ultra-low overpotentials of 0.31 and 0.33 V, respectively. Finally, spin magnetic moment of active center is found to be a catalytic descriptor for ORR on the DACs. Our work will stimulate the experimental exploration of the ultra-high-density DACs and provides the novel insight into the relationship between ORR activity of the DACs and their spin states.

---

[*] Corresponding author. E-mail: madw@henu.edu.cn, dwmachina@126.com (DW. Ma); zslu@htu.edu.cn (ZS. Lu)

## I. INTRODUCTION

The four-electron oxygen reduction reaction (ORR) is a key electrochemical reaction for the renewable energy conversion and storage technologies, due to its significance for metal-air batteries and proton exchange membrane fuel cells (PEMFCs). [1–3] Up to now, Pt-based catalysts are the best-known ORR electrocatalysts for commercial applications. [4–6] However, the large-scale commercialization of Pt-based catalysts is significantly restricted by the low natural reserves, high cost, and limited stability of Pt. Therefore, designing and searching alternative ORR electrocatalysts of low cost, high activity, and long durability is increasingly attractive but with great challenges ahead. [7,8] For example, single-atom catalyst (SAC) with uniform dispersion of transition-metal (TM) active sites coordinated with nitrogen atoms in carbon (termed as M-N-C, such as $FeN_4$ SAC [9,10]) have gained wide attentions in the past few years. [11–14]

Very recently, double-atom catalyst (DAC) has emerged as a new frontier in heterogenous electrocatalysis due to its synergetic dual atomic sites, which can endow DAC with many intrinsic advantages compared with SAC for the multiple-step coupled electron-proton transfer reactions in electrocatalysis. [15–17] To be specific, for ORR, the dual atomic sites can facilitate the dissociation of the O-O bond of the intermediates tending to adopt the side-on adsorption configuration, which favors the desirable four-electron ORR and hinders the two-electron ORR, thereby promoting the energy efficiency of the ORR process in PEMFCs or metal-air batteries and enhancing the stability of catalysts. [18–23,23–30] Among the various DACs for ORR, the one with each TM metal coordinated with four N atoms in carbon sheet (termed as $M_2N_6$ DAC as shown in **Fig. S1(a)** within the Supplemental Material [31]) has gained particular attentions, [18,23–29,32,33] for which the basic structural motif, $M_2N_6$, is shown in **Fig. 1(a)**. Combining experimental and theoretical simulations, the $FeMnN_6$, $FeCoN_6$, $FeNiN_6$, and $FeZnN_6$ DACs for ORR have been thoroughly investigated, and these systems outperform the corresponding SAC counterparts and even the Pt/C catalysts. [23–26,28,29] For example, electronic synergies between Fe and Mn in $FeMnN_6$ deliver the better durability and more excellent ORR performance (half-wave potentials are 0.928 and 0.804 V in alkaline condition and acidic media, respectively) than both $FeN_4$, $MnN_4$ SACs, and commercial Pt/C catalyst. [24] In addition, besides ORR, $M_2N_6$ DACs have been also theoretically and experimentally demonstrated to be efficient for the electrocatalytic $CO_2$ reduction and other reactions. [34–39]

On the other hand, one of the most significant challenges facing the atomically dispersed catalysts is their ultra-low active site density (typically less than 5 wt%), [23–25,27,34,39–43] which leads to their overall poor catalytic performance and limits the future industrialization. For example, for $MnFeN_6$ DAC mentioned above, the content of Fe and Mn are only 2.3 wt% and 1.6 wt%, respectively. [24] Excitingly, very recently, there already are some breakthroughs in synthesizing ultra-high-density (UHD) SACs. [44–48] A multilayer stabilization strategy was used to construct SACs with metal loading to 16 wt%. [49] A versatile approach combining impregnation and two-step annealing can successfully synthesize 15 metals on chemically distinct supports with metal contents up to 23 wt%. [50] The meal loading of SAC can even reach up to about 40 wt% by means of a graphene quantum dot assisted synthesis strategy. [51] However, with advantages of DAC compared with SAC in mind, to our best knowledge, there are no UHD-DACs reported up to now, which is the focus of the present study.

Herein, starting from the $M_2N_6$ motif shown in **Fig. 1(a)**, we tried to construct the UHD-DACs (**Fig. 1(a)**) with as few C atoms as possible as glue, which can reach the metal loading > 40 wt% in weight fraction, corresponding to the 14 at% in atomic metal percentage. For the active center, all the 3$d$ TMs (Sc~Zn) with small atomic radius have been considered, inspired by the experimental reports. [24,26,34,35] Among various UHD-DAC candidates, ten ones were screened out, which exhibit excellent dynamic, thermal, mechanical, thermodynamic, and electrochemical stabilities, comparable to the corresponding low-density (LD) systems (see the atomic models in **Fig. S1(b)**). Furthermore, the ORR reaction mechanism has been comparably investigated for the UHD-DACs and their corresponding LD-DACs (**Fig. S1(b)**). Finally, the relationship between the ORR catalytic activity of the considered DACs and their spin magnetic moment was investigated and uncovered.

## II. COMPUTATIONAL DETAILS

All spin-polarized computations were carried out by the generalized gradient approximation (GGA) method with Perdewe–Burke–Ernzerhof (PBE) functional [52] based on density functional theory (DFT) implemented in the Vienna *ab initio* Simulation Package (VASP), [53,54] in which van der Waals (vdW) correction proposed by Grimme (DFT+D3) was chosen. [55] The plane-wave basis set with a cut-off energy of 500 eV were employed. A vacuum layer of ~16 Å was used to avoid the interactions between periodic images for all calculations. The convergence thresholds of the total energy and the Hellmann–Feynman force are $10^{-7}$ eV and 0.0005 eV/Å, respectively. For the structural optimization of primitive cell, various magnetic states have been considered and relaxed by the Monkhorst–Pack meshes of 8 × 8 × 1 and the optimized lattice parameters of the lowest-energy structures are presented in **Table S1**, with the corresponding atomic configurations in **Fig. S1(c)**. The phonon dispersions were calculated with the finite displacement method by using the PHONOPY code. [56] The *ab initio* molecular dynamics (AIMD) simulations for the new UHD-DAC supercells (2 × 2 × 1) were performed based on the NVT ensemble with a time step of 2 fs and total time of 10 ps. The simulated scanning tunneling microscopy (STM) images were obtained using the Tersoff–Hamann theory. [57]

For the binding energy ($E_b$), dissolution potential ($U_{diss}$, versus SHE), and formation energy ($E_{form}$) of UHD-DACs and corresponding LD-DACs, they can be defined as:

$$E_b = E(\text{total}) - E(\text{CN}) - \mu(M_1) - \mu(M_2) \tag{1}$$

$$U_{\text{diss}} = U_{\text{diss}}°(\text{metal, bulk}) - E_b/(eN_e \ast N_M) \tag{2}$$

$$E_{\text{form}} = E(\text{total}) - 6\mu(N) - 6\mu(C) - \mu(M_1) - \mu(M_2) \tag{3}$$

where $E(\text{total})$ and $E(\text{CN})$ are the total energies of DAC system and CN composite, respectively; $\mu(M_1)$, $\mu(M_2)$, $\mu(N)$, and $\mu(C)$ are the chemical potential of the involved species, which is taken from the metal bulk, $N_2$ molecules, and graphene, respectively; $U_{\text{diss}}°(\text{metal, bulk})$ and $N_e$ are the standard dissolution potential (pH = 0) of bulk metal in aqueous solution and the number of electrons involved in the dissolution, respectively, which are taken from the previous work [58,59] and has been listed in **Table S4**. $N_M$ indicates the number of metal atoms. The $N_M$ takes 2 for the current DAC system. [60]

The in-plane Young's modulus ($E_x$, $E_y$), Poisson's ratio ($\nu_{xy}$, $\nu_{yx}$), and shear modulus ($G$) along the armchair ($x$) and zigzag ($y$) directions for UHD-DACs can be calculated by the following equations: [61]

$$E_x = (C_{11}C_{22} - C_{12}C_{21})/C_{22} \tag{4}$$

$$E_y = (C_{11}C_{22} - C_{12}C_{21})/C_{11} \tag{5}$$

$$\nu_{xy} = C_{21}/C_{22} \tag{6}$$

$$\nu_{yx} = C_{12}/C_{11} \tag{7}$$

$$G = C_{66} \tag{8}$$

For the ORR related calculations, the standard conventional cells of various UHD-DACs were used as the electrocatalysts and the Monkhorst–Pack meshes of 5 × 4 × 1 and 12 × 10 × 1 are adopted for structural optimization and calculation of densities of states (DOS), respectively. The convergence thresholds of the total energy and the Hellmann–Feynman force are $10^{-5}$ eV and 0.03 eV/Å, respectively.

The free energy change ($\Delta G$) under the acidic medium (pH = 0) for each elementary reaction step was calculated based on the computational hydrogen electrode (CHE) model, [62–64] according to the following equation:

$$\Delta G = \Delta E + \Delta E_{\text{ZPE}} - T\Delta S \tag{9}$$

where $\Delta E$ is the reaction energy from DFT calculations. $\Delta E_{\text{ZPE}}$ and $T\Delta S$ ($T$ = 298.15 K) are the contributions of the zero-point energy and entropy to $\Delta G$, respectively. $E_{\text{ZPE}}$ and $TS$ for the free molecules are taken from the NIST database, [65] and those of the adsorbed species were obtained based on the calculated vibrational frequencies and then with the VASPKIT code. [66] The reaction step with $\Delta G_{\text{max}}$ is the potential-determining step (PDS).

The ORR has been investigated under the acidic medium (pH = 0), the corresponding elementary reaction steps along the pathway-A, pathway-B, and pathway-C for ORR in **Fig. 3a** can be described as:

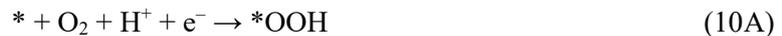

$$\ast + O_2 + H^+ + e^- \rightarrow \ast OOH \tag{10A}$$

$$* + O_2 + H^+ + e^- \rightarrow *O*OH \tag{10B; 10C}$$

$$*OOH + H^+ + e^- \rightarrow *O + H_2O \tag{11A}$$

$$*O*OH + H^+ + e^- \rightarrow *O + H_2O \tag{11B}$$

$$*O*OH + H^+ + e^- \rightarrow *OH*OH \tag{11C}$$

$$*O + H^+ + e^- \rightarrow *OH \tag{12A; 12B}$$

$$*OH*OH + H^+ + e^- \rightarrow *OH + H_2O \tag{12C}$$

$$*OH + H^+ + e^- \rightarrow * + H_2O \tag{13A; 13B; 13C}$$

The Gibbs adsorption free energy of the intermediates (*OH, *OOH (*O*OH), *OH*OH, and *O) can be evaluated by the following equations:

$$\Delta G\,(*OH) = G\,(*OH) + 1/2\,G\,(H_2) - G* - G\,(H_2O) \tag{14}$$

$$\Delta G\,(*OOH) = G\,(*OOH) + 3/2\,G\,(H_2) - G* - 2G\,(H_2O) \tag{15}$$

$$\Delta G\,(*O*OH) = G\,(*O*OH) + 3/2\,G\,(H_2) - G* - 2G\,(H_2O) \tag{16}$$

$$\Delta G\,(*OH*OH) = G\,(*OH*OH) + G\,(H_2) - G* - 2G\,(H_2O) \tag{17}$$

$$\Delta G\,(*O) = G\,(*O) + G\,(H_2) - G* - * G\,(H_2O) \tag{18}$$

According to the above $\Delta G$ values, and setting $\Delta G\,(O_2) = 4.92$ eV and $\Delta G\,(H_2O) = 0$ eV, respectively, we can obtain the free energy changes ($\Delta G_x$, for step $x = 1-4$) and equilibrium potential ($U_x$, $x = 1-4$) as follows:

$$\Delta G_1 = -eU_1 = \Delta G\,(*OOH) - 4.92 \tag{19A}$$

$$\Delta G_1 = -eU_1 = \Delta G\,(*O*OH) - 4.92 \tag{19B; 19C}$$

$$\Delta G_2 = -eU_2 = \Delta G\,(*O) - \Delta G\,(*OOH) \tag{20A}$$

$$\Delta G_2 = -eU_2 = \Delta G\,(*O) - \Delta G\,(*O*OH) \tag{20B}$$

$$\Delta G_2 = -eU_2 = \Delta G\,(*OH*OH) - \Delta G\,(*O*OH) \tag{20C}$$

$$\Delta G_3 = -eU_3 = \Delta G\,(*OH) - \Delta G\,(*O) \tag{21A; 21B}$$

$$\Delta G_3 = -eU_3 = \Delta G\,(*OH) - \Delta G\,(*OH*OH) \tag{21C}$$

$$\Delta G_4 = -eU_4 = -\Delta G\,(*OH) \tag{22A; 22B; 22C}$$

where A, B and C indicate the pathway-A, pathway-B, and pathway-C for ORR in **Fig. 3a**, respectively. Then the overpotential ($\eta_{ORR}$, V) is defined in the following:

$$\eta_{ORR} = (\Delta G_{max}/e) + 1.23 \tag{23}$$

The standard GGA simulations usually underestimates the solvation effect to affect the free energies for ORR under the practical conditions. The continuum solvation model [67] were used to investigate the solvation effect on the ORR. The solvation energies for different adsorbed species are shown in the **Table S6**. These values from continuum solvation model are all negative, indicating that the solvent can stabilize the adsorbates in solution. Also, it is noted that our calculation values are comparable to the previous works. [18,59,68] We also studied the solvation effect for ORR from the explicit water by combining force-field molecular dynamic simulation implemented in LAMMPS, [69,70] *ab initio* molecular dynamics simulation in VASPsol, [67] and static structural optimization. It is found that the solvation energies from the continuum solvation model by VASPsol are comparable with those from our explicit model and those of previous work. [18,68,71] Moreover, the explicit solvation effect doesn't alter their optimal reaction pathway for ORR on both UHD- and LD-DACs [see Note S1 with corresponding figures and tables in Supplemental Material [31]].

## III. RESULTS AND DISCUSSION

### A. Morphologies of UHD-DACs

Stimulated by the experimental LD-DAC (**Fig. S1(a)**), we have designed a new kind of 2D material (*i.e.*, the UHD-DAC herein in **Fig. 1(a)**) by extremely reducing the surrounding carbon atoms of dual metal active center. From **Fig. 1(a)**, the lattice structure of $M_2N_6$ UHD-DACs was constructed by gluing the $M_2N_6$ motif by six C atom locating at the corner of primitive cell, which realize a minimum cell containing a pair of TM atoms (TM = Sc~Zn). Thus, the geometric structure of active center for the designed UHD-DACs is the same as that of experimentally prepared LD-DAC, but the former achieves a unprecedently ultra-high density of active center with metal loading up to 36 to 46 wt% for DACs in weight fraction, which catch up with the recently reported ultra-high-density Ir-N-C SAC (~ 40 wt%). [51] From the view of atomic metal percentage, [72] here the UHD-DACs present the high metal contents of 14.3 at%, about 4 times larger than that of the Ir-N-C SAC (3.8 at%). [51] For the structural optimization, the optimized lattice parameters of the lowest-energy structures are presented in **Table S1**, with the corresponding atomic configurations in **Fig. S1(c)**. For all the cases, the systems have the 2D rhombus lattice of primitive cell, while they own the 2D rectangular conventional cell, dissimilar to the well-known lattice symmetry of graphene. The homonuclear (heteronuclear) UHD-DAC structures possess the space group of *Cmmm* (*Amm*2) with the point group $D_{2h}$ ($C_{2v}$) symmetry, which both belong to the orthorhombic phase. As a contrast, the theoretical model of experimentally synthesized LD-DACs was also built by embedding the same $M_2N_6$ active center in 3 × 6 rectangular supercells (**Fig. S1(b)**) for simulations.

### B. Dynamic, thermal, mechanical, thermodynamic, and electrochemical stabilities of UHD-DACs

We comprehensively examined their stabilities from five aspects: dynamic, thermal, mechanical, thermodynamic, and electrochemical stabilities (**Fig. 1(b)**). The constructed UHD-DAC would be considered stable if it meets all the stability criteria (**Fig. S2**). First, the dynamic stability of the homonuclear $M_2N_6$ UHD-DACs was verified by calculating their phonon dispersions, which describes the dispersion relations of lattice vibrations and usually identifies as the decisive factor for materials' stability. From **Fig. 2(a)** and **Fig. S3**, we can find that only homonuclear $Mn_2N_6$, $Fe_2N_6$, $Co_2N_6$, and $Ni_2N_6$ UHD-DACs are dynamically stable as indicated by the absence of the obvious imaginary phonon frequencies. Small atomic radii of Mn, Fe, Co, and Ni correspond to the relatively small lattice constants and compact lattice interactions [73] of UHD-DACs, which are response for the dynamic stability of homonuclear $Mn_2N_6$, $Fe_2N_6$, $Co_2N_6$, and $Ni_2N_6$ UHD-DACs. Based on this result, it is expected that the heteronuclear UHD-DACs formed by the pairwise combination of these four elements also have the good dynamic stability. Thus, we also investigated the phonon dispersions of six heteronuclear $M_2N_6$ UHD-DACs, *i.e.*, $MnFeN_6$, $MnCoN_6$, $MnNiN_6$, $FeCoN_6$, $FeNiN_6$, and $CoNiN_6$ (**Fig. S3**). Like their corresponding homonuclear ones, they also possess the great dynamic stability without any imaginary frequencies. It is noticed that all the UHD-DAC structures have 52 phonon vibrational modes, which are 3 acoustic modes and 49 optical modes. Through group theory analysis, the irreducible representation of the optical branches at the Brillouin-zone with Raman (R) and infrared (IR) symbols for the homonuclear and heteronuclear UHD-DAC structures are as follows: $\Gamma_{optic}$ (homonuclear) = $7A_g$ (R) + $2A_u$ + $7B_{1g}$ (R) + $4B_{1u}$ (IR) + $3B_{2g}$ (R) + $6B_{2u}$ (IR) + $4B_{3g}$ (R) + $6B_{3u}$ (IR); $\Gamma_{optic}$ (heteronuclear) = $13A_1$ (R/IR) + $6A_2$ (R) + $7B_1$ (R/IR) + $13B_2$ (R/IR). We hope here the Raman and IR Brillouin spectroscopy measurements in the further experiments will test and verify the accuracy of the calculations in our theoretical modeling.

Furthermore, *ab initio* Molecular Dynamics (AIMD) simulations were performed to check the thermal stability under ambient conditions of these 10 $M_2N_6$ UHD-DACs with dynamic stability. Note that their total free energy only shows the periodic oscillating near the equilibrium state during the entire simulation period of 10 ps at 500 K (**Fig. 2(b)** and **Fig. S4**). Simulation snapshots of the $M_2N_6$ UHD-DACs also show their structural integrity at 10 ps, implying their thermal stability. In addition, we calculated their elastic constants (**Table S3**) of $M_2N_6$ UHD-DACs using the finite differences method [74] with assuming a thickness of 3.34 Å on the basis of graphene's thickness [75] to check their mechanical stability. Significantly, these values satisfy the requirements of the mechanical stability criterion [76,77] of a 2D material, i.e., $C_{11} \times C_{22} - C_{12}^2 > 0$ and $C_{66} > 0$. The in-plane Young's modulus ($Y$, GPa), Poisson's ratio ($v$), and shear modulus ($G$, GPa) were obtained from the elastic constants, which are summarized in **Table S3**. Also, the 2D directional projection in polar coordinates of the Young's modulus, Poisson's ratio, and shear modulus of $Fe_2N_6$ UHD-DAC are displayed in **Fig. 2(c)**, and those for other systems are illustrated in **Fig. S5**. The mechanical results in **Fig. 2(c)**, **Fig. S5**, and **Table S3** show that the in-plane Young's modulus and Poisson's ratio along the armchair direction are larger than that along the zigzag direction, revealing the moderate anisotropy of the mechanical properties. Also, it is noted that the Young's modulus of all the UHD-DACs are much smaller than that of graphene (~1000 GPa), [75,78] indicating that these UHD-DACs are softer than the well-known graphene and suitable for the strain-tunable catalysis. [79]

The thermodynamic stability of ten UHD-DACs was examined by the binding energy ($E_b$), as shown in **Fig. 2(d)** and **Table S4**. The more negative $E_b$ values indicates the stronger chemical bond formed between the metal atoms and the coordinated N atoms. We can see that, the $E_b$ values for all the ten UHD-DACs are more negative than -5 eV, which can effectively prevent the diffusion and aggregation of the metal atoms, promising high thermodynamic stability of the whole systems. For comparison, the $E_b$ for the corresponding LD-DACs were also calculated, and the $E_b$ values of UHD-DACs are only slightly positive than those of the corresponding LD-DACs, suggesting that the high density of metal loading will not affect the thermodynamic stability of current DAC systems.

The electrochemical stability of electrocatalyst is one of the key factors for its practical application in an electrochemical environment. [59] According to its definition in the Computational Details, the more positive dissolution potential ($U_{diss}$) suggests that the metal dimer strongly bind with the coordinated atoms in the UHD-DAC system and the dissolution of metal atoms can be avoided under the acidic conditions (pH = 0). To this end, we calculated the $U_{diss}$ of above ten UHD-DAC systems, as illustrated in **Fig. 2(e)** and **Table S4**. Here we use the range of $U_{diss}$ > 0 V to measure the electrochemical stability, which is a valid evaluation criterion that has been widely used in the literatures. [59,80] Given that all ten DAC systems have the positive $U_{diss}$ values, the metal dimers in the corresponding UHD-DACs ($Mn_2N_6$, $Fe_2N_6$, $Co_2N_6$, $Ni_2N_6$, $MnFeN_6$, $MnCoN_6$, $MnNiN_6$, $FeCoN_6$, $FeNiN_6$, and $CoNiN_6$) with the dynamic, thermal, mechanical, and thermodynamic stability can survive under the experimentally electrochemical conditions, suggesting their excellent electrochemical stability.

**C. Synthesis feasibility and electronic properties of UHD-DACs**

Then the feasibility of experimental realization of these ten UHD-DACs is explored by calculating the formation energies ($E_{form}$) following in Computational Details. As shown in **Fig. 2(f)** and **Table S4**, their $E_{form}$ values range from about 1.7 to 3.1 eV and are generally smaller than those of another kinds of $M_2N_6$ catalysts with double $MN_3$ groups. [81,82] We also observed that the formation energies of $Ni_2N_6$, $FeNiN_6$, and $CoNiN_6$ UHD-DACs are lower than or comparable with those of the corresponding synthesized ones with low metal contents. [23,28,29,39,43] Noted that the values of $E_{form}$ values depend on the chemical potentials of the involved species, which in turn depends on the experimental conditions. Therefore, it is the relative values of $E_{form}$ that determine the formation probability of a specific system among various ones and it is also expected that all the proposed UHD-DAC systems with dynamic, thermal, mechanical, thermodynamic, and electrochemical stabilities are suitably synthesizable in the experiments. Moreover, the calculated $E_{form}$ values of UHD-DACs also indicate the thermodynamical stability of N-N motif and suppression of $N_2$ gas formation in the UHD-DACs because all these values are lower than that the experimentally synthesized $Ni_2N_6$ LD-DAC (3.22 eV). [39] Hence, we also theoretically predicted the STM images of these ten stable UHD-DACs with 8 × 8 supercells for the future experimental identification, as shown in **Fig. 2(g)** and **Fig. S6**. It is easy to recognize and correlate them with the corresponding atomic structure of metal dimers, while the C and N atoms are difficult to identify because they are lighter and then brighter than the metal ones.

We further investigated the electronic properties of these ten stable UHD-DACs, including the electron localization function (ELF) and density of states (DOS). The bonding characters can be effectively characterized by ELF. As shown in **Fig. S7**, the electrons are more localized around C and N atoms with larger ELF values, while more delocalized around metal atoms with smaller ELF values, indicating the strong covalent bonding characteristics for C-C, C-N, N-N, and Fe-N interactions and metallic bonding features of metal dimers, which is responsible for the good stability of the UHD-DACs. Importantly, the calculated total DOS (TDOS) and partial DOS (PDOS) based on PBE and HSE06 functional [83] presented in **Fig. S8-9**, suggest the metallic conductivity feature of all stable UHD-DACs, beneficial for the charge transfer during the electrocatalytic ORR process. Finally, for most of the systems there are large spin magnetic moments localized on the anchored metal atoms, which benefit the effective adsorption and activation of the oxygenated intermediates. [84–86] However, the spin magnetic moments of the embedded Ni atoms are fully quenched in the $Ni_2N_6$ and $CoNiN_6$ UHD-DACs, probably due to the charge transfer and the electronic state coupling.

**D. Scaling relationship for adsorption of key intermediates**

Following the comprehensive assessment of the overall stabilities and electronic properties of ten UHD-DACs, we investigated their ORR catalytic activity, as well as that of the corresponding LD-DACs. Considering the dual metal sites of DACs, three reaction pathways for ORR have been considered under acidic conditions (pH = 0). The traditional ORR pathway over the catalysts is shown in **Fig. 3(a)**: the protonation of *OOH to produce the first $H_2O$ molecule and the continuous hydrogenation of the remaining *O atom to yield the second $H_2O$ molecule (pathway-A). Moreover, the dual-metal sites may facilitate the breaking of the O-O bond of *OOH as described in **Fig. 3(a)** to drive the ORR following the pathway-B or pathway-C. [19,87] All the three pathways have been considered for ORR to obtain the most energetically feasible ones.

In fact, there are two mechanisms for the first electron transfer process of ORR. [88] One is that the process of short-range electron transfer to adsorbed $O_2$ occurs in the inner Helmholtz plane (ET-IHP mechanism), for which $O_2$ protonates in the catalyst surface to form the *OOH. Another is that the process of long-range electron transfer to non-adsorbed $O_2$ occurs in the outer Helmholtz plane (ET-OHP mechanism), for which $O_2$ protonates in the electrolytes and subsequently adsorbs at active site as *OOH. However, the ET-IHP mechanism usually occurs on Pt based noble metal catalyst. For the SACs and DACs, the ET-OHP mechanism is usually considered to be the relevant one. [11,26] Thus, we focused on the latter ET-OHP mechanism that does not rely on direct adsorption of $O_2$ to the catalyst surface and the initial stage for ORR is the gas phase $O_2$.

Firstly, we investigated the adsorption of key intermediates, including *O, *OOH, *O*OH,

*OH*OH, and *OH, and the scaling relationship between their binding strengths (*O and *OOH versus *OH (pathway-A); *O and *O*OH versus *OH (pathway-B); *OH*OH and *O*OH versus *OH (pathway-C)). From **Fig. 3(b)-(d)**, we can see that for UHD-DACs pathway-A exhibits an excellent linear scaling relationship between the binding strengths of the oxygenated intermediates with $R^2$ = 0.99 and 0.98 for $\Delta G$ (*O) and $\Delta G$ (*OOH) versus $\Delta G$ (*OH), respectively. For pathway-B, the binding strength of *O and *OH is also well correlated with $R^2$ = 0.98. However, the binding strengths between *O*OH or *OH*OH and *OH for pathway-B and C display relatively poor linear relationship (with $R^2$ < 0.71), which could help to achieve better ORR activity due to the deviation of the linear scaling relationship. Moreover, this poor scaling relation can be ascribed to the different adsorption modes of *O*OH and *OH*OH, both of which bind with the active sites through two metal atoms, compared with *OH binding through one metal atom (**Fig. S10**). The flexible dual-atom active sites induced decoupling binding strength between the key intermediates have been observed in other reactions, such as electrocatalytic $CO_2$ reduction and $N_2$ reduction. [89–91] Note that for the LD-DAC systems, comparatively, the scaling relationship between the binding strengths of key intermediates (**Fig. S11**) have the similar linear trends with the above UHD-DACs.

**E. ORR catalytic activity of UHD-DACs**

For the three ORR pathways, the reaction free energy diagrams on UHD-DACs under different potentials are shown in **Fig. S12-S14**, and the theoretical overpotentials ($\eta_{ORR}$) for the most favorable pathway are summarized in **Fig. 3(e)**. We can observe that the homonuclear $Mn_2N_6$, $Fe_2N_6$, and $Co_2N_6$ UHD-DACs prefers to the pathway-C, in which the O-O bond of *OOH is broken and *OH*OH species is formed with each O atom binding with one metal atom. The heteronuclear UHD-DACs (except $CoNiN_6$ with pathway-A) tend to adopt the pathway-B, where the dissociated *OOH, i.e., *O*OH, is reduced to *O. Among them, $Fe_2N_6$ and $Co_2N_6$ UHD-DACs have the smallest $\eta_{ORR}$ of 0.31 and 0.33 V along the optimal pathway-C, respectively, delivering the highest ORR activity among all the UHD-DACs. Note that the $\eta_{ORR}$ of $Fe_2N_6$ and $Co_2N_6$ UHD-DACs are smaller than that for the state-art commercial Pt/C catalyst (0.45 V). [1,92] Moreover, interestingly, the Fe- and Co-based heteronuclear UHD-DACs also have the relatively good catalytic activity, in which the $MnFeN_6$, $MnCoN_6$, $FeCoN_6$, $FeNiN_6$, and $CoNiN_6$ UHD-DACs possess the $\eta_{ORR}$ of 0.66, 0.68, 0.51, 0.52, and 0.59 V, respectively. For comparison, we also investigated the ORR activity over the corresponding LD-DACs, for which the free energy diagrams are presented in **Fig. S15-17**, and the $\eta_{ORR}$ of the optimal pathways are also summarized in **Fig. 3(e)**. We can see the $\eta_{ORR}$ for all the UHD-DACs (except $Co_2N_6$) are comparable with those of the corresponding LD-DACs, indicating that the UHD-DACs can well maintain the intrinsic ORR activity of the active centers. Interestingly, the $Co_2N_6$ UHD-DAC even exhibits much higher ORR activity than its LD-DAC case. More significantly, FeMnN$_6$, FeCoN$_6$, FeNiN$_6$, and FeZnN$_6$ DACs have been definitely determined to exhibit excellent catalytic performances for the ORR. [23–26,28,29] Overall, the excellent stabilities and ORR catalytic activities render the proposed UHD-DACs promising ORR electrocatalysts with ultra-high-density active sites available.

Moreover, we considered the competitive reactions on the UHD-DACs, including the two-electron ORR and hydrogen evolution reaction (HER). The results show that most of the concerned UHD-DACs have the good selectivity for four-electron ORR toward H$_2$O (see the details in Supplemental Material [31] with **Table S5**).

To better understand the ORR catalytic trend, the $\eta_{ORR}$ as a function of $\Delta G$ (*OH) are plotted in **Fig. 4**. It can be found that for the pathway-A (**Fig. 4(a)**), pathway-B (**Fig. 4(b)**), and pathway-C (**Fig. 4(c)**), when $\Delta G$ (*OH) reaches about 0.9, 1.0, and 1.0 eV, respectively, the DAC systems possess the highest ORR catalytic activity (the lowest $\eta_{ORR}$), and the weaker or stronger binding of *OH on the active sites will lead to the deteriorated ORR activity. Consequently, a volcanic relationship between $\eta_{ORR}$ and $\Delta G$ (*OH) can be observed for all the reaction pathways. Thus, $\Delta G$ (*OH) can serve as an effective ORR activity descriptor for the studied DAC systems. Moreover, for the optimal reaction pathways of each DAC (**Fig. 4(d)**), the appropriate binding strength of *OH on Fe$_2$N$_6$ and Co$_2$N$_6$ UHD-DACs prompts the equilibrium between oxygenated species activation and catalyst recovery and then results in the highest activity among all the UHD-DACs.

## F. Spin as universal descriptor for both UHD- and LD-DACs

Recent studies show that the spin magnetic moment of the active center is a novel activity descriptor for the ORR on SACs, which is an intrinsic physical property. [93] For example, the larger spin magnetic moments can induce a better catalytic activity for ORR on the FeN$_4$ based SACs because the spin electrons are beneficial for binding and activation of reaction intermediates. [94] Correspondingly, we investigated the possibility of the spin magnetic moment as ORR activity descriptor herein. Interestingly, as shown in **Fig. 5(a)**, $\Delta G$ (*OH) for all UHD- and LD-DAC systems are correlated linearly with spin magnetic moments $M_S$ of the metal atom that binds OH ($R^2 = 0.87$), which shows that the larger spin magnetic moment contributes to the stronger adsorption of *OH. This strong linear correlation suggests that $M_S$ can function as an efficient descriptor to predict the ORR catalytic activity. As expected, as shown in **Fig. 5(b)**, the $\eta_{ORR}$ indeed exhibits a volcano relationship with $M_S$, and importantly $M_S$ is a unified descriptor for both UHD- and LD-DAC systems.

Importantly, the excellent linear relationship between $\Delta G$ (*OH) and $M_S$, and volcano relationship between $\Delta G$ (*OH) and $\eta_{ORR}$, have also been confirmed by the HSE06 functional (**Fig. S19**). From **Fig. S19(a)** and **Fig. S19(c)**, there is a good linear scaling relationship between $M_S$ and $\Delta G$ (*OH) both for PBE and HSE methods, with the linear slope of -0.47 and -0.35, respectively. The relatively smaller $M_S$ for magnetic active center by PBE method is responsible for the more negative slope, compared with that from HSE method. Moreover, there is a good volcanic relationship for $\eta_{ORR}$ as a function of $M_S$ for PBE and HSE calculations (**Fig. S19(b)** and **Fig. S19(d)**). It is found that the critical $M_S$ corresponding to the theoretical optimal activity by PBE (0.94 $\mu_B$) is smaller than that based on HSE functional (1.32 $\mu_B$), which also stems from the smaller $M_S$ for magnetic active center by PBE method. These results suggest that the spin related conclusions are reliable from PBE method, and as one of the key points in our work, spin magnetic moment of active center by PBE functional indeed can act as a catalytic descriptor for ORR on the DACs, similar to that by HSE functional.

Finally, we explored the mechanism for the significantly enhanced ORR activity of $Co_2N_6$ UHD-DAC compared with its LD-DAC counterpart, for which the free energy diagrams are presented in **Figs. 6(a)** and **6(b)**. From above, we know that excellent ORR activity of $Co_2N_6$ UHD-DAC results from its proper binding of *OH, which correlates with the spin magnetic moment of the active center. Therefore, we comparatively studied the spin magnetic states of $Co_2N_6$ UHD-DAC and LD-DAC. As presented in **Figs. 6(c)** and **6(d)**, the PDOS show that the Co 3$d$ orbitals are asymmetric in UHD-DAC system while symmetric in LD-DAC system, indicating the (non)magnetic states of Co atoms for (LD-) UHD-DAC. Moreover, the Co $d_{x2}$, $d_{xz}$, and $d_{yz}$ in-plane orbitals mainly contribute to the spin magnetic moment of the Co atom in UHD-DAC.

Based on the crystal field theory, the Co cation in the $CoN_4$ square planar crystal field for SAC system possesses the 3$d^7$ electronic configuration, [95] which should have the unpaired electrons and give rise to the spin magnetic state. However, in the current DAC systems, there are double $CoN_4$ square planar crystal field interacting each other. For the $Co_2N_6$ LD-DAC, the short Co-Co distance (2.25 Å) can intensively share the $d$ orbital electrons, resulting in the faultlessly pairing of in-plane $d$ orbital electrons ($d_{x2}$, $d_{xz}$, and $d_{yz}$) and the completely quenching of spin magnetic moment in LD-DAC. In contrast, a large Co-Co distance (2.45 Å) in UHD-DAC contributes to the retention of the spin magnetic moment. The spin density distribution in the illustrations of **Figs. 6(c)** and **6(d)** can also confirm our above analysis about the difference of spin magnetic states between $Co_2N_6$ UHD- and LD-DACs.

In addition, we studied the electronic state interaction between Co 3$d$ and *OH 2$sp$, for which the PDOS are presented in **Fig. S20**, and furthermore a quantitative analysis in **Figs. 6(e)** and **6(f)**, resorting to the crystal orbital Hamilton populations (COHP). For $Co_2N_6$ UHD-DAC, the adsorption of *OH almost leads to the complete spin quenching of the Co atom that binds it (see **Fig. S20**), and thus the spin-up and spin-down bands contribute similar binding strengths with the integrated COHP (ICOHP) of -1.23 and -1.16 eV, respectively. On the contrary, for $Co_2N_6$ LD-DAC, the adsorption of *OH can induce spin magnetic moment on the Co atoms. Significantly, spin-up bands contribute the similar binding strength, compared with the case of

$Co_2N_6$ UHD-DAC, with ICOHP of -1.20 eV, while much more occupation of the antibonding state for spin-down bands leads to the much smaller contribution to *OH binding with ICOHP of -0.85 eV. Consequently, $Co_2N_6$ UHD-DAC binds *OH much stronger than $Co_2N_6$ UHD-DAC, and exhibits high ORR activity.

## IV. CONCLUSIONS

In summary, we theoretically confirmed the existence of ultra-high-density DACs with metal loading > 40 wt%, constructed from the experimentally synthesized $M_2N_6$ motif. Among the investigated systems, ten ones ($Mn_2N_6$, $Fe_2N_6$, $Co_2N_6$, $Ni_2N_6$, $MnFeN_6$, $MnCoN_6$, $MnNiN_6$, $FeCoN_6$, $FeNiN_6$, and $CoNiN_6$) were demonstrated to have well dynamic, thermal, mechanical, thermodynamic, and electrochemical stabilities. Mechanism studies on ORR show that most of the UHD-DACs have comparable ORR activities with the corresponding LD-DACs, and $Fe_2N_6$ and $Co_2N_6$ UHD-DACs locate at the peak of the activity volcano with the ultra-low overpotentials of 0.31 and 0.33 V, respectively. Furthermore, we investigated the relationship between the ORR activities and the spin states of the active centers, and identified that the spin magnetic moment of active centers of the DACs can serve as an effective catalytic descriptor. Interestingly, $Co_2N_6$ UHD-DACs exhibit significantly enhanced ORR activity compared with the $Co_2N_6$ LD counterpart, which can be ascribed to the distortion of the square planar crystal field induced spin state crossover. In addition, the simulated STM images and symmetry classifications of phonon modes for the UHD-DAC systems provide a basis for the future experimental confirmation. We hope our prediction will simulate the experimental exploration of ultra-high-density DACs for practical catalytic applications, and the identified activity descriptor, spin magnetic moment, will guide the rational design of efficient DACs based on spin-state regulation.

**ACKNOWLEDGEMENTS**

This work is supported by the National Natural Science Foundation of China (Grant No. 12204151, 12274118), the Program for Science & Technology Innovation Talents in Universities of Henan Province (Grant No. 20HASTIT028), the Special Project for Fundamental Research in University of Henan Province (No. 22ZX013), the China Postdoctoral Science Foundation (Grant No. 2022M711048), and the Open Project Program of Guangdong Provincial Key Laboratory of Electronic Functional Materials and Devices, Huizhou University (EFMD2022009M). The work was carried out at National Supercomputer Center in Tianjin, and this research was supported by TianHe Qingsuo Project-special fund project.

Maps, and DOS of Various UHD-DACs; Atomic Configurations of *OOH, *O*OH, *OH*OH, *O, and *OH Species on $Fe_2N_6$ and $MnFeN_6$ UHD-DACs; Scaling Relationship of Adsorption for Key Intermediates; Free Energy Diagrams of ORR for All Stable UHD- and LD-DACs for Pathway-A, B, and C; PDOS with Electronic Orbital Interaction for $Co_2N_6$ UHD- and LD-DACs with *OH Adsorption; Selectivity Analysis; Solvation Effect Analysis with Solvation Energies for Different Adsorbed Species, which includes Refs. [18, 58-59, 67-71].

# Figures

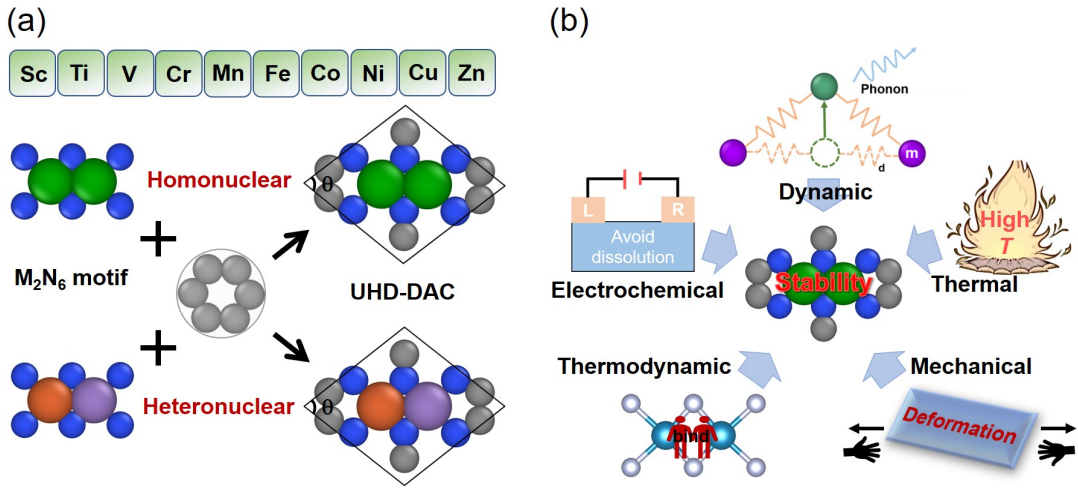

**FIG. 1.** (a) The schematic depiction of $M_2N_6$ UHD-DACs construction (M = Sc, Ti, V, Cr, Mn, Fe, Co, Ni, Cu, and Zn) from experimental $M_2N_6$ motif with the bonding of carbon atoms, including the homonuclear and heteronuclear DACs. (b) Five considered stabilities for $M_2N_6$ UHD-DACs: dynamic, thermal, mechanical, thermodynamic, and electrochemical stabilities.

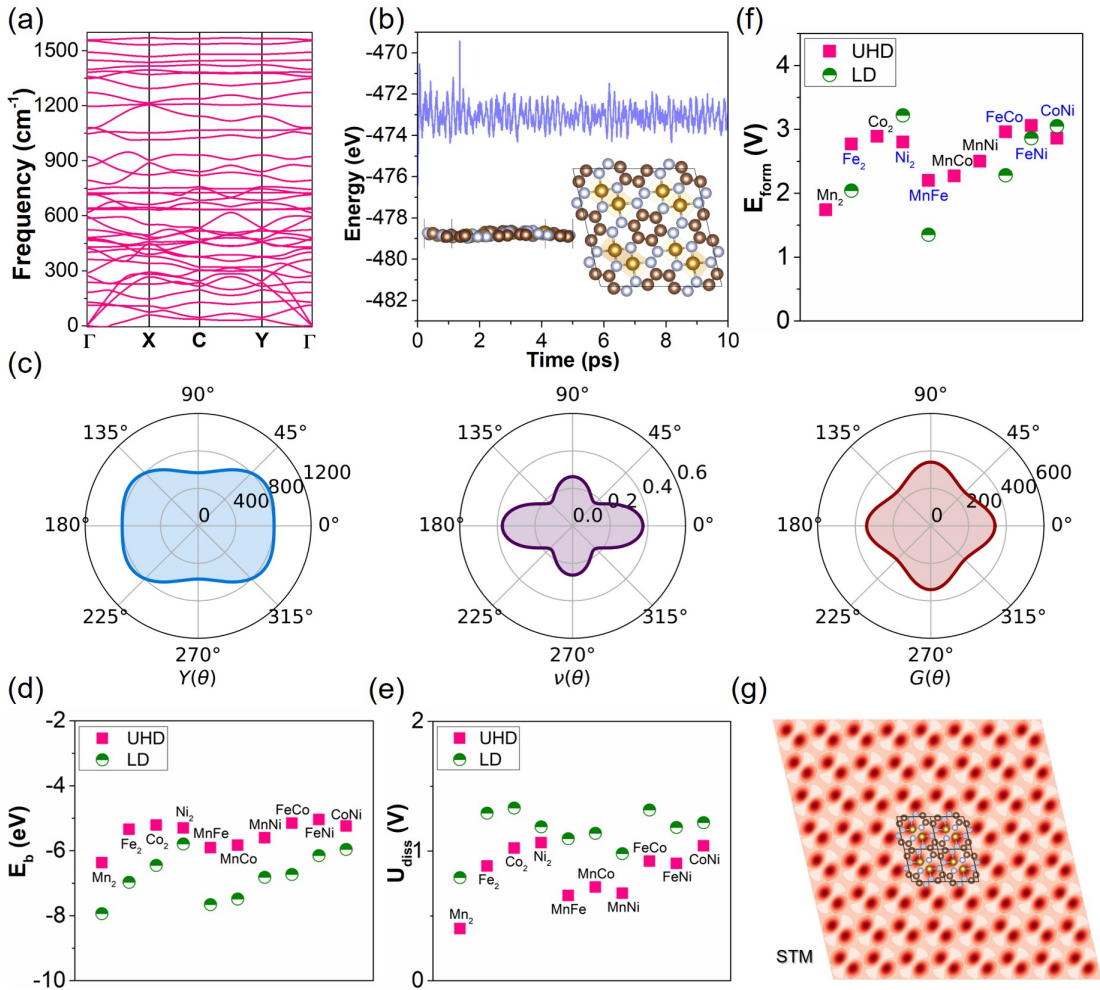

**FIG. 2.** (a) Phonon dispersions and (b) total free energy fluctuation during the AIMD

simulations for $Fe_2N_6$ UHD-DAC. In (b), The AIMD simulation was performed under 500 K for 10 ps with a time-step of 2 fs. The side and top views of the atomic configuration at 10 ps are shown as insets. (c) The 2D projection in polar coordinates of the Young's modulus ($Y$, GPa), Poisson's ratio ($v$), and shear modulus ($G$) of $Fe_2N_6$ UHD-DAC. (d) Binding energies ($E_b$) and (e) dissolution potentials ($U_{diss}$) of double metal atoms for ten considerable $M_2N_6$ UHD-DACs with dynamic stability and their corresponding LD-DACs. (f) Formation energies ($E_{form}$) for ten considerable $M_2N_6$ UHD-DACs with dynamic stability and some corresponding experimental LD-DACs. (g) The simulated STM image for $Fe_2N_6$ UHD-DAC with the bias voltage of 1.0 V for the size of for 8 × 8 supercells. The tip was considered to be separated from the sample by a vacuum barrier width of 3.5 Å.

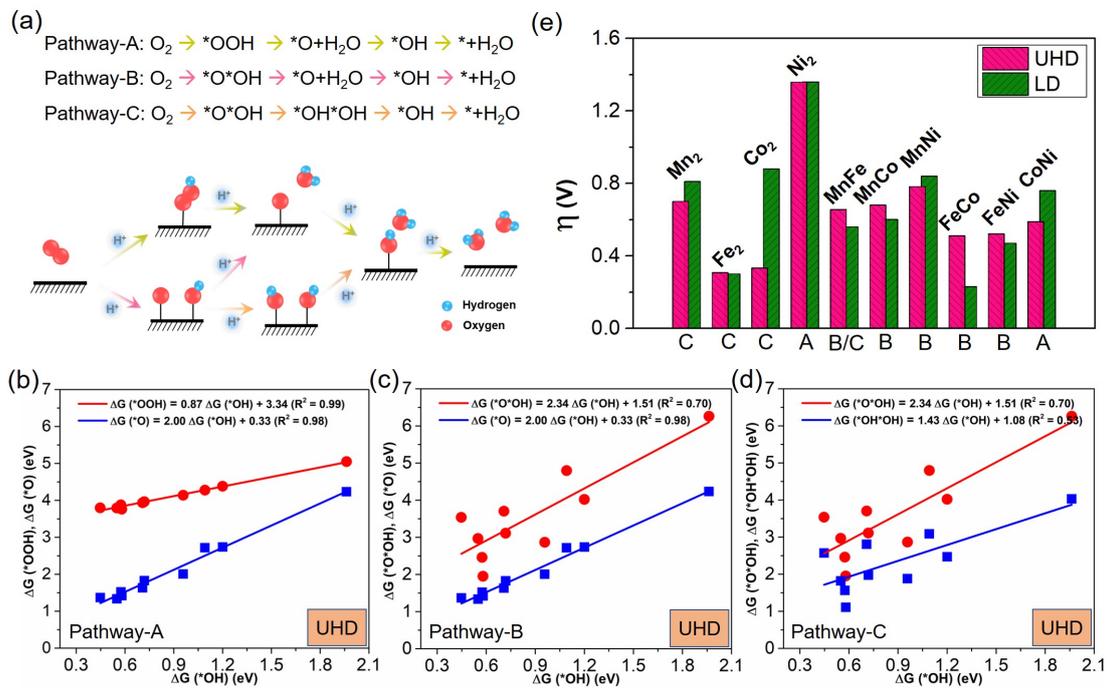

**FIG. 3.** (a) The schematic depiction of the three possible reaction pathways for the ORR at pH = 0. The scaling relationship between the Gibbs adsorption free energy of the oxygenated intermediates for (b) pathway-A, (c) pathway-B, and (d) pathway-C over the corresponding UHD-DACs. (e) The overpotentials ($\eta_{ORR}$) of ORR through the optimal reaction pathway (A, B or C) over corresponding UHD- and LD-DACs.

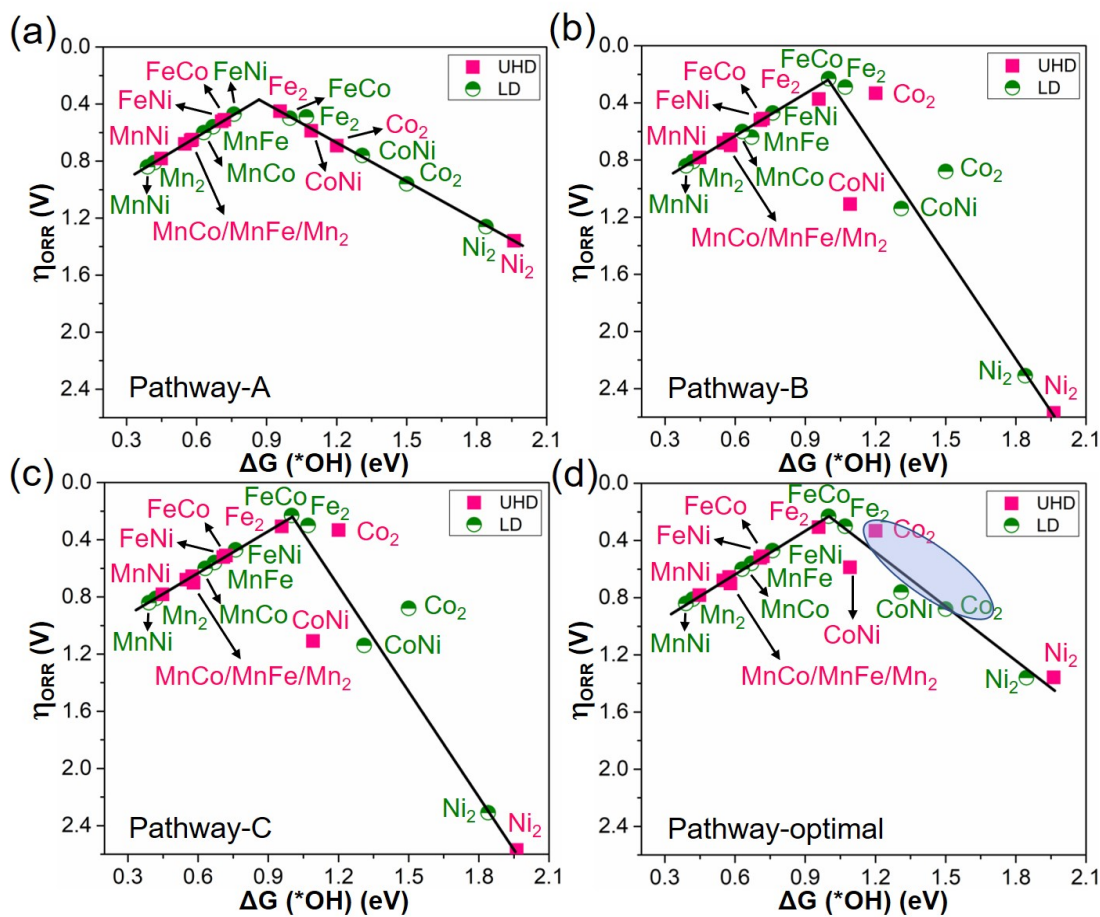

**FIG. 4.** Volcano plots for the $\eta_{ORR}$ as a function of the Gibbs adsorption free energy $\Delta G$ (*OH) for (a) pathway-A, (b) pathway-B, (c) pathway-C, and (d) optimal pathway. The pink and green marks represent the UHD- and LD-DACs, respectively.

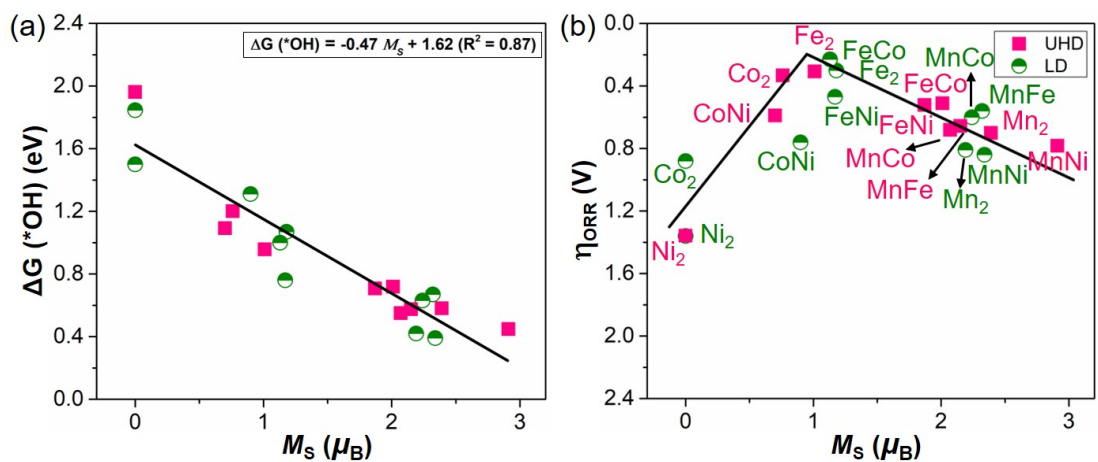

**FIG. 5.** (a) The scaling relationship between the Gibbs adsorption free energy of *OH ($\Delta G$ (*OH)) in optimal pathway and the local spin magnetic moments ($M_S$) for the metal atom that anchor the *OH. (b) The volcano plots for the $\eta_{ORR}$ for optimal pathway as a function of the local spin magnetic moments for the metal atom that anchor the *OH. The pink and green marks represent the UHD- and LD-DACs, respectively.

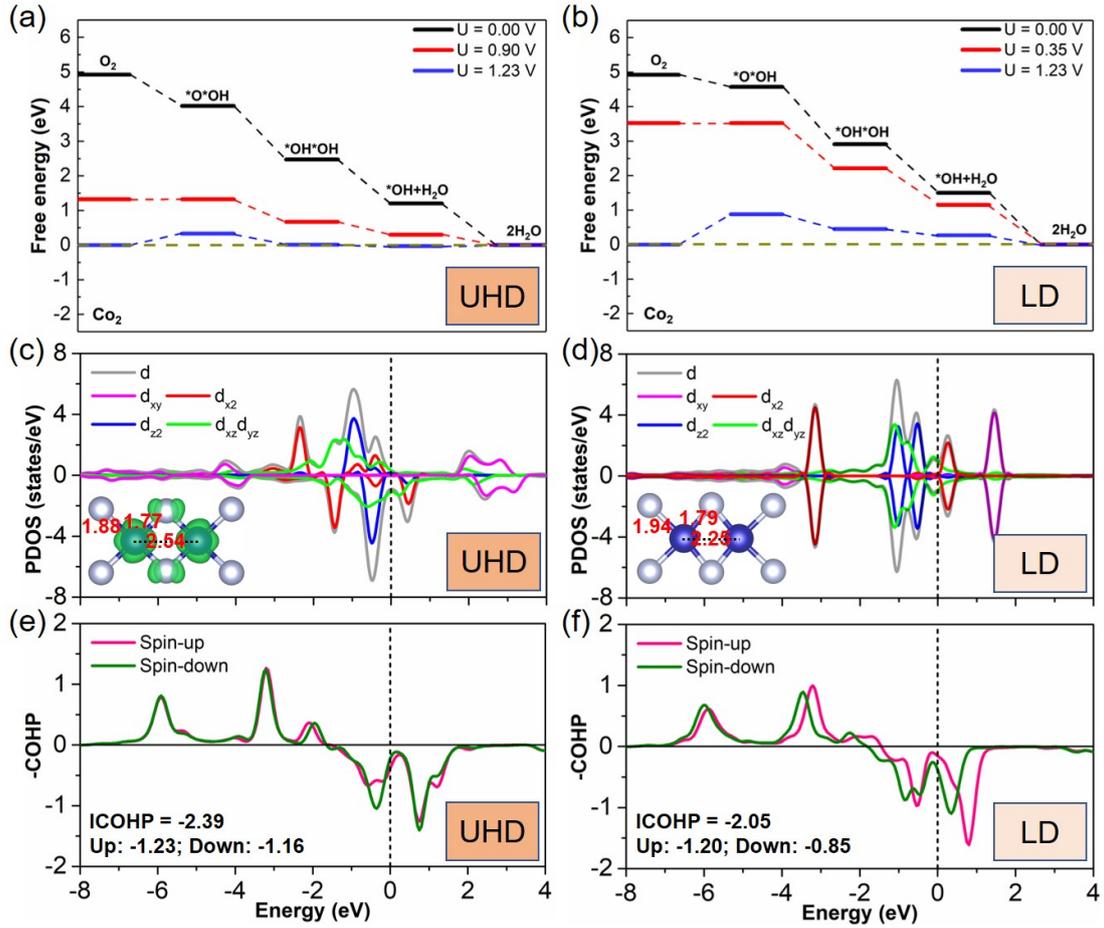

**FIG. 6.** The reaction free energy diagrams of ORR for $Co_2N_6$ (a) UHD- and (b) LD-DACs at different potentials for optimal pathway-C. The PDOS of Co 3$d$ orbitals for $Co_2N_6$ (c) UHD- and (d) LD-DACs without *OH adsorption. Here the Fermi level was set to 0. The insert figure is the corresponding spin density (0.005 e/bohr$^3$) for $Co_2N_6$ UHD- and LD-DACs. The COHP for the interaction between Co and *OH for (e) UHD- and (f) LD-DACs.

# Supplemental Material for
# "Ultra-high-density double-atom catalyst with spin moment as activity descriptor for oxygen reduction reaction"

Peng Lv[1], Wenjing Lv[1], Donghai Wu[1], Gang Tang[2], Xunwang Yan[3], Zhansheng Lu[4*], Dongwei Ma[1*]

[1]*Key Laboratory for Special Functional Materials of Ministry of Education, and School of Materials Science and Engineering, Henan University, Kaifeng 475004, China*

[2]*Advanced Research Institute of Multidisciplinary Science, Beijing Institute of Technology, Beijing 100081, China*

[3]*College of Physics and Engineering, Qufu Normal University, Qufu 273165, China*

[4]*School of Physics, Henan Normal University, Xinxiang 453007, China*

---

[*] Corresponding author. E-mail: madw@henu.edu.cn, dwmachina@126.com (DW. Ma); zslu@htu.edu.cn (ZS. Lu)

**Table S1.** The structural parameters of UHD-DACs, including the lattice constants $a$ (=$b$), the spin stable state (SSS) of ferromagnetic (FM), antiferromagnetic (AFM), ferrimagnetic (FiM), and nonmagnetic (NM), and the theoretical metal loading (ML, wt%).

| UHD-DACs | $a$ (Å) | $\theta$ (Å) | SSS | ML |
|---|---|---|---|---|
| $Sc_2$ | 6.89 | 105.2 | NM | 36.5 |
| $Ti_2$ | 6.75 | 104.7 | FM | 38.0 |
| $V_2$ | 6.64 | 103.8 | FM | 39.5 |
| $Cr_2$ | 6.57 | 103.0 | AFM | 40.0 |
| $Mn_2$ | 6.53 | 103.5 | AFM | 41.3 |
| $Fe_2$ | 6.46 | 103.1 | FM | 41.7 |
| $Co_2$ | 6.48 | 104.9 | FM | 43.0 |
| $Ni_2$ | 6.51 | 106.3 | NM | 42.9 |
| $Cu_2$ | 6.59 | 106.1 | FM | 44.9 |
| $Zn_2$ | 6.70 | 106.0 | NM | 45.6 |
| MnFe | 6.49 | 103.3 | FiM | 41.5 |
| MnCo | 6.48 | 103.4 | FM | 42.2 |
| MnNi | 6.54 | 105.3 | FM | 42.1 |
| FeCo | 6.50 | 104.7 | FM | 42.4 |
| FeNi | 6.51 | 105.2 | FM | 42.3 |
| CoNi | 6.51 | 106.3 | FM | 43.0 |

**Table S2.** The spin magnetic moments ($\mu_B$) of M1 and M2 in the UHD- and LD-DACs. The values without (with) parentheses are calculated from PBE (HSE06) functional. Due to the large calculations for HSE06 functional, we only focus on the $Co_2N_6$ DAC among the LD-DAC system.

|  | ULD-DACs |  | LD-DACs |  |
| --- | --- | --- | --- | --- |
| System | M1 | M2 | M1 | M2 |
| $Mn_2$ | 2.36 (3.20) | -2.36 (-3.20) | 2.39 | -2.39 |
| $Fe_2$ | 1.03 (1.68) | 1.03 (1.68) | 1.11 | 1.11 |
| $Co_2$ | 0.75 (1.14) | 0.75 (1.14) | 0.00 (0.00) | 0.00 (0.00) |
| $Ni_2$ | 0.00 (0.00) | 0.00 (0.00) | 0.00 | 0.00 |
| MnFe | 2.33 (3.14) | -0.83 (-2.18) | 2.32 | -1.19 |
| MnCo | 2.09 (3.22) | -0.04 (-0.73) | 2.24 | -0.04 |
| MnNi | 2.92 (3.43) | 0.05 (0.01) | 2.22 | 0.01 |
| FeCo | 2.00 (2.39) | 0.52 (1.02) | 1.13 | -0.02 |
| FeNi | 1.86 (2.41) | 0.08 (0.02) | 2.18 | 0.16 |
| CoNi | 0.69 (1.03) | 0.00 (0.00) | 0.90 | 0.17 |

**Table S3.** The mechanical properties of UHD-DACs, including the elastic constants $C_{ij}$ (GPa), in-plane Young's modulus $E_x$ and $E_y$ (GPa), in-plane Poisson's ratio $v_{xy}$ and $v_{yx}$, in-plane shear modulus $G$ (GPa).

| $M_2$ | $C_{11}$ | $C_{22}$ | $C_{12}$ | $C_{66}$ | $E_x$ | $E_y$ | $v_{xy}$ | $v_{yx}$ | $G$ |
|---|---|---|---|---|---|---|---|---|---|
| $Mn_2$ | 894.61 | 629.46 | 235.31 | 340.41 | 806.65 | 567.57 | 0.37 | 0.26 | 340.41 |
| $Fe_2$ | 929.58 | 712.57 | 232.14 | 341.13 | 853.95 | 654.60 | 0.33 | 0.25 | 341.13 |
| $Co_2$ | 850.87 | 673.34 | 293.51 | 363.02 | 722.93 | 572.10 | 0.44 | 0.34 | 363.02 |
| $Ni_2$ | 829.60 | 733.99 | 264.77 | 365.75 | 734.10 | 649.49 | 0.36 | 0.32 | 365.75 |
| MnFe | 915.74 | 674.40 | 234.16 | 343.23 | 834.44 | 614.52 | 0.35 | 0.26 | 343.23 |
| MnCo | 915.69 | 688.81 | 242.73 | 339.40 | 830.15 | 624.47 | 0.35 | 0.27 | 339.40 |
| MnNi | 810.35 | 668.02 | 289.44 | 351.09 | 684.94 | 564.64 | 0.43 | 0.36 | 351.09 |
| FeCo | 851.50 | 653.41 | 293.94 | 357.65 | 719.27 | 551.94 | 0.45 | 0.35 | 357.65 |
| FeNi | 820.89 | 655.66 | 300.26 | 358.37 | 683.38 | 545.83 | 0.46 | 0.37 | 358.37 |
| CoNi | 828.65 | 735.04 | 262.32 | 365.08 | 735.03 | 652.00 | 0.36 | 0.32 | 365.08 |

**Table S4.** The binding energy ($E_b$), dissolution potential ($U_{diss}$, versus SHE), and formation energy ($E_{form}$) for considered UHD-DACs and their corresponding LD-DACs. It should be noted that this definition for $E_b$ takes the metal bulk as the reference system while $E_b'$ uses the free metal atom as reference. Experimental values for $U_{diss}°$ (pH = 0) and $N_e$ are taken from reference. [58,59] The $E_{form}$ values marked by red color for LD-DACs are the experimentally synthesized systems.

| DACs | $E_b$ | | $E_b'$ | | $U_{diss}°$ | $N_e$ | $U_{diss}$ | | $E_{form}$ | |
|---|---|---|---|---|---|---|---|---|---|---|
| | UHD | LD | UHD | LD | | | UHD | LD | UHD | LD |
| Mn$_2$ | -6.37 | -7.94 | -14.65 | -16.22 | -1.19 | 2 | 0.40 | 0.79 | 1.74 | 1.07 |
| Fe$_2$ | -5.34 | -6.97 | -16.28 | -17.91 | -0.45 | 2 | 0.88 | 1.29 | 2.77 | 2.04 |
| Co$_2$ | -5.21 | -6.45 | -16.47 | -17.71 | -0.28 | 2 | 1.02 | 1.33 | 2.89 | 2.56 |
| Ni$_2$ | -5.30 | -5.79 | -15.90 | -16.39 | -0.26 | 2 | 1.07 | 1.19 | 2.80 | 3.22 |
| MnFe | -5.91 | -7.66 | -15.52 | -17.27 | -0.82 | 2 | 0.66 | 1.09 | 2.20 | 1.35 |
| MnCo | -5.83 | -7.49 | -15.60 | -17.26 | -0.74 | 2 | 0.72 | 1.14 | 2.27 | 1.52 |
| MnNi | -5.60 | -6.82 | -15.04 | -16.26 | -0.73 | 2 | 0.67 | 0.98 | 2.50 | 2.19 |
| FeCo | -5.15 | -6.73 | -16.25 | -17.83 | -0.37 | 2 | 0.92 | 1.32 | 2.96 | 2.28 |
| FeNi | -5.04 | -6.15 | -15.81 | -16.92 | -0.36 | 2 | 0.90 | 1.18 | 3.06 | 2.86 |
| CoNi | -5.24 | -5.96 | -16.17 | -16.89 | -0.27 | 2 | 1.04 | 1.22 | 2.86 | 3.05 |

**Table S5.** The $\Delta G$ (*O), $\Delta G_1$ (ORR), and $\Delta G$ (*H) on ten stable UHD-DACs.

| UHD-DACs | $\Delta G$ (*O) | $\Delta G_1$ (ORR) | $\Delta G$ (*H) |
|---|---|---|---|
| $Mn_2$ | 1.42 | -1.16 | 0.32 |
| $Fe_2$ | 2.01 | -0.78 | 0.32 |
| $Co_2$ | 2.74 | -0.54 | 0.48 |
| $Ni_2$ | 4.23 | 0.13 | 0.37 |
| MnFe | 1.52 | -1.04 | 0.37 |
| MnCo | 1.34 | -1.13 | 0.33 |
| MnNi | 1.37 | -1.12 | 0.43 |
| FeCo | 1.83 | -0.96 | 0.30 |
| FeNi | 1.63 | -0.98 | 0.38 |
| CoNi | 2.72 | -0.64 | 0.31 |

Herein, we have investigated the competitive reactions on the UHD-DACs, including the two-electron ORR and hydrogen evolution reaction (HER). For competitive two-electron ORR toward $H_2O_2$ production, the Gibbs free energy of *O ($\Delta G$ (*O)) for a catalyst with a high selectivity should be more positive than 3.52 eV. [59] Our results showed that the $\Delta G$ (*O) for most of the UHD-DACs are less than 3.52 eV, indicating their poor selectivity for two-electron ORR toward $H_2O_2$. Only one exception, *i.e.*, $Ni_2N_6$ UHD-DAC, has the larger $\Delta G$ (*O) with 4.23 eV, which indeed exhibits high two-electron ORR selectivity. Herein, we consider the four-electron ORR toward $H_2O$ on $Ni_2N_6$ UHD-DAC with the aim to investigate the activity descriptor for all DAC systems. For competitive HER, we calculated the *H adsorption on various UHD-DACs and the comparison results of $\Delta G$ (*H) and $\Delta G_1$ (ORR) (here $\Delta G_1 = \Delta G$ (*OOH) – 4.92)) are organized in **Table S5**. We found that all the UHD-DACs have the positive $\Delta G$ (*H) values but negative $\Delta G_1$ (ORR) values, indicating that they have a strong ability to suppress the competing HER.

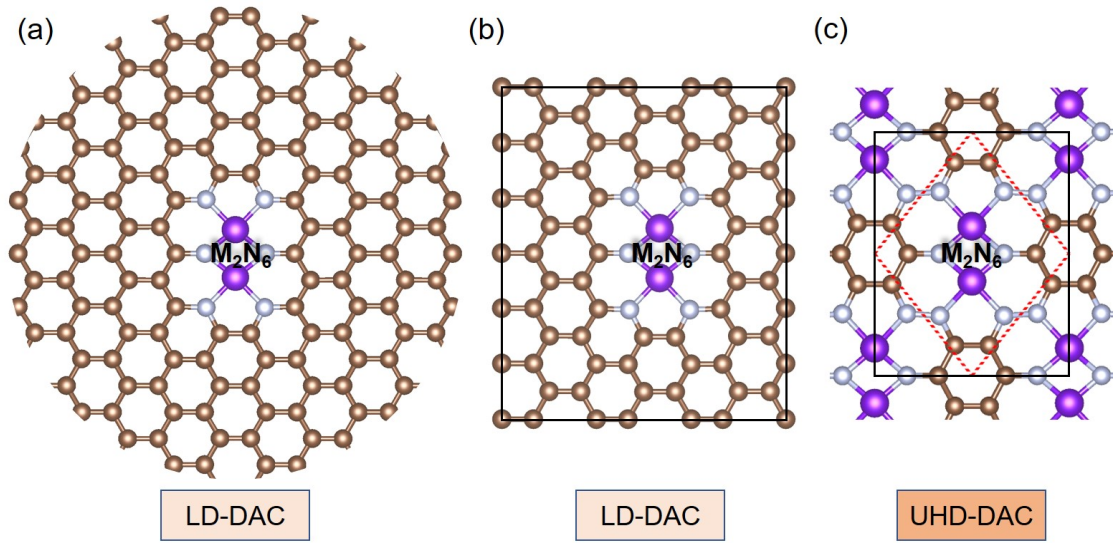

**Fig. S1.** (a) The atomic structure for experimental $M_2N_6$ LD-DAC with each transition-metal atom coordinated with four N atoms in carbon sheet ($M_2N_6$ with two $MN_4$ motifs). (b) For comparison with UHD-DAC, the theoretical model of experimentally synthesized LD-DACs with low metal contents is built by embedding the same $M_2N_6$ active center in 3 × 6 rectangular supercells. (c) The atomic structure for our constructed UHD-DAC. The red dash lines imply the primitive cell and the black solid lines denote the conventional cell. We can find our constructed UHD-DAC and experimentally prepared LD-DAC both have the same active center, *i.e.*, metal dimers with the same N coordination ($M_2N_6$ with two $MN_4$ motifs). The difference is that they contain different amounts of carbon in their substrate, resulting in different metal contents and densities of active site.

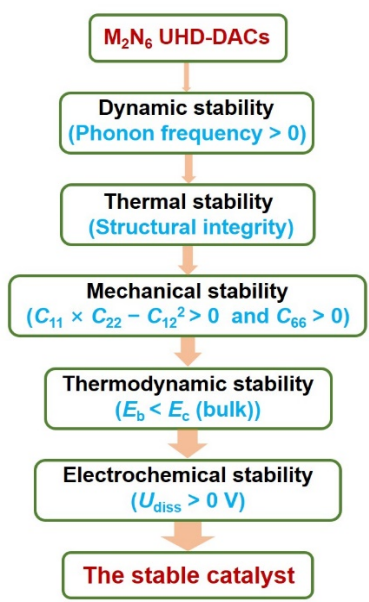

**Fig. S2**. The screening strategy with limiting criterion for the selection of stable UHD-DACs, including five aspects: dynamic, thermal, mechanical, thermodynamic, and electrochemical stabilities (**Fig. 1(b)**).

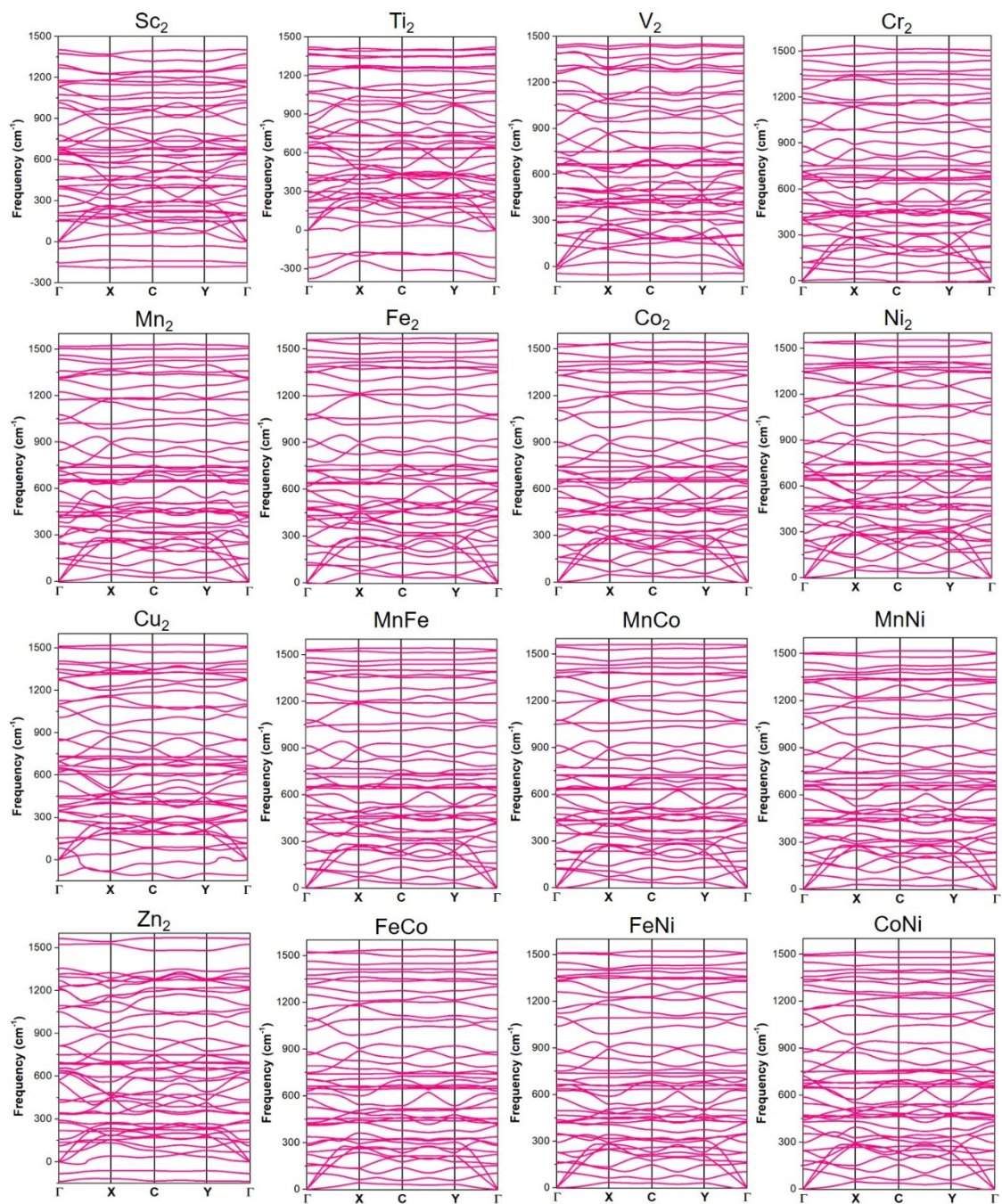

**Fig. S3.** The phonon dispersions of various UHD-DACs. The $Fe_2N_6$ case have also been shown in the **Fig. 2** of main text.

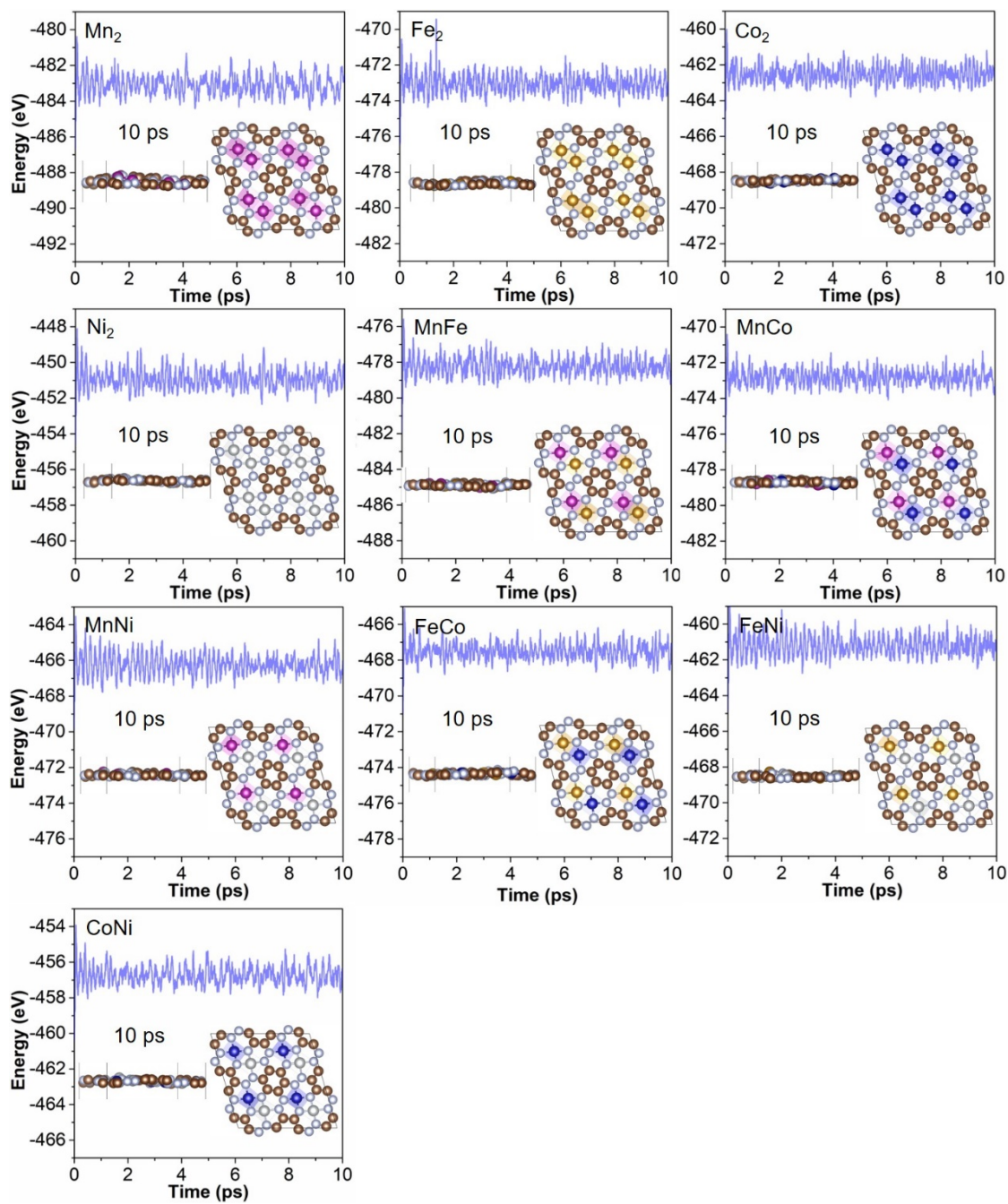

**Fig. S4.** The total free energy fluctuations of various UHD-DACs during the AIMD simulations at 500 K for 10 ps. The Fe$_2$N$_6$ case have also been shown in the **Fig. 2** of main text.

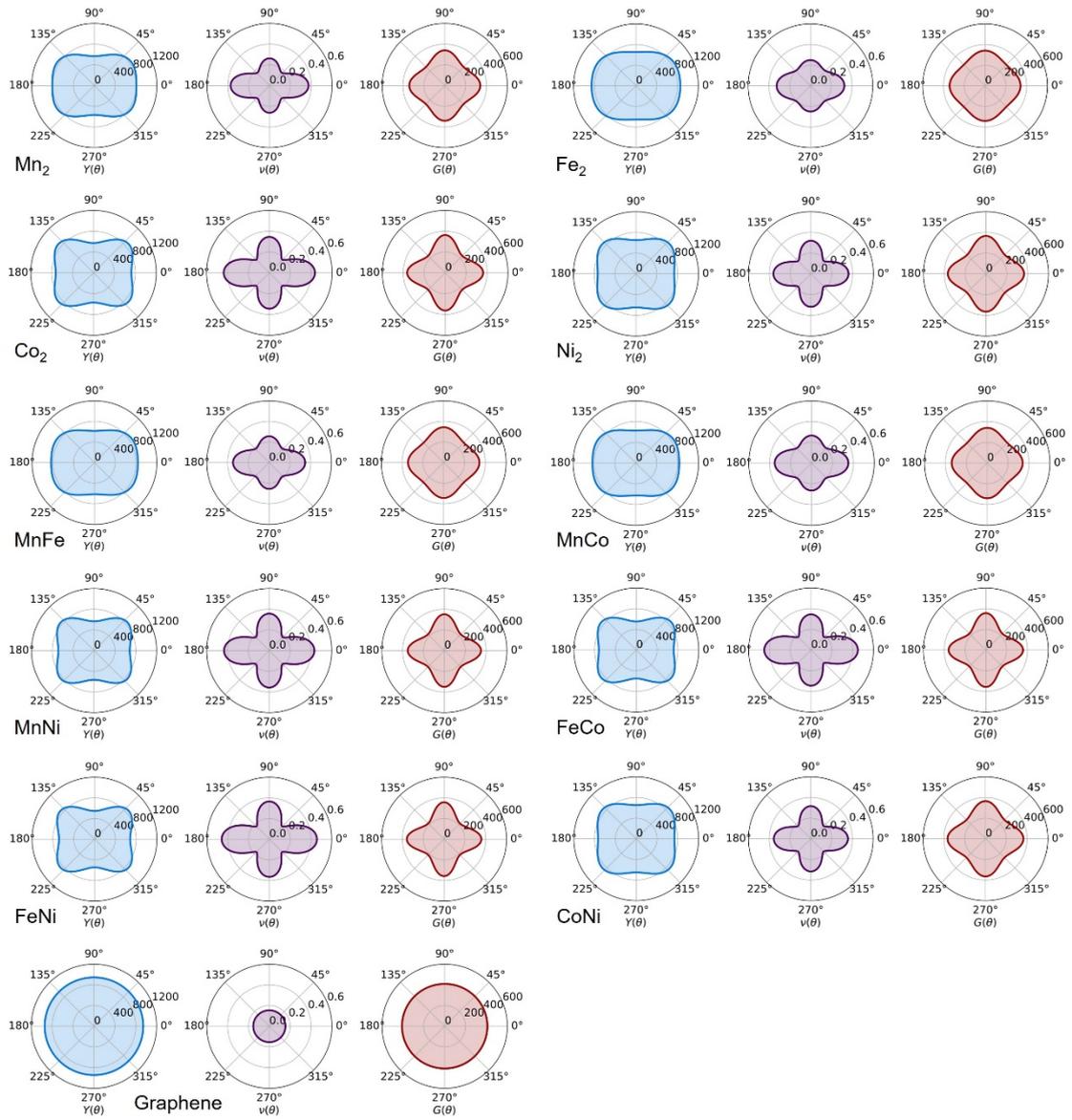

**Fig. S5.** The 2D projection in polar coordinates of the Young's modulus ($Y$, GPa), Poisson's ratio ($v$), and shear modulus ($G$) of various UHD-DACs. The mechanical properties of graphene are also calculated for the comparison. The $Fe_2N_6$ case have also been shown in the **Fig. 2** of main text.

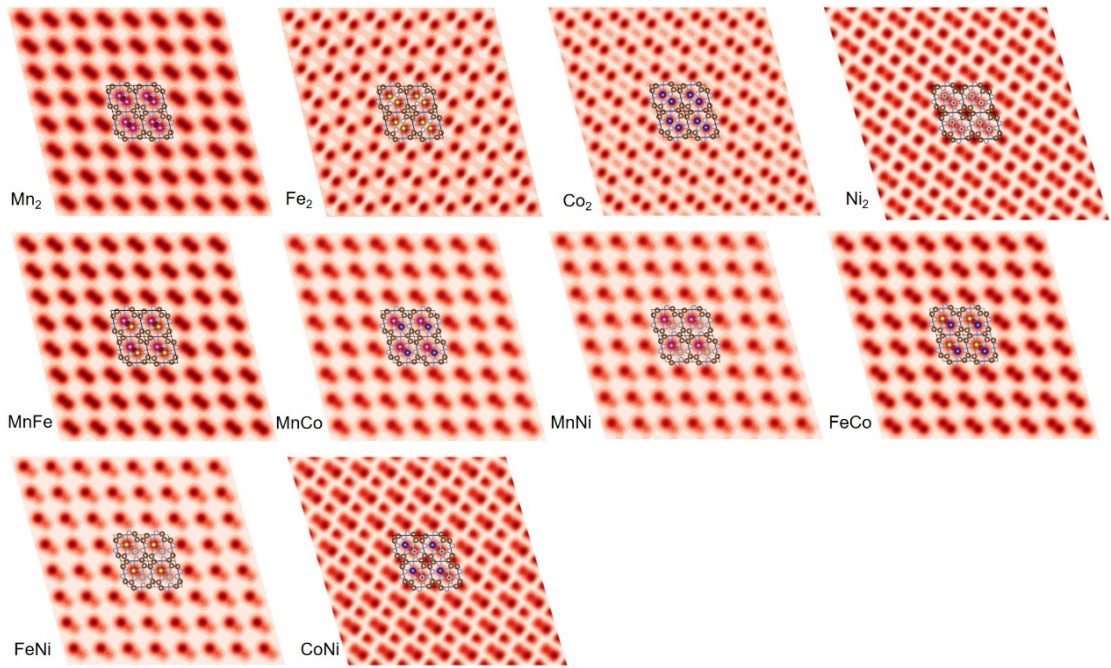

**Fig. S6.** The simulated STM images for the size of for 8 × 8 supercells with the different bias voltage (1.0 V for $Mn_2N_6$, $Fe_2N_6$, $Co_2N_6$, $Ni_2N_6$, and $CoNiN_6$ UHD-DACs, 2.0 V for $MnFeN_6$, $MnCoN_6$, and $FeCoN_6$ UHD-DACs, 3.0 V for $MnNiN_6$ and $FeNiN_6$ UHD-DACs). The tip was considered to be separated from the sample by a vacuum barrier width of 3.5 Å. The $Fe_2N_6$ case have also been shown in the **Fig. 2** of main text.

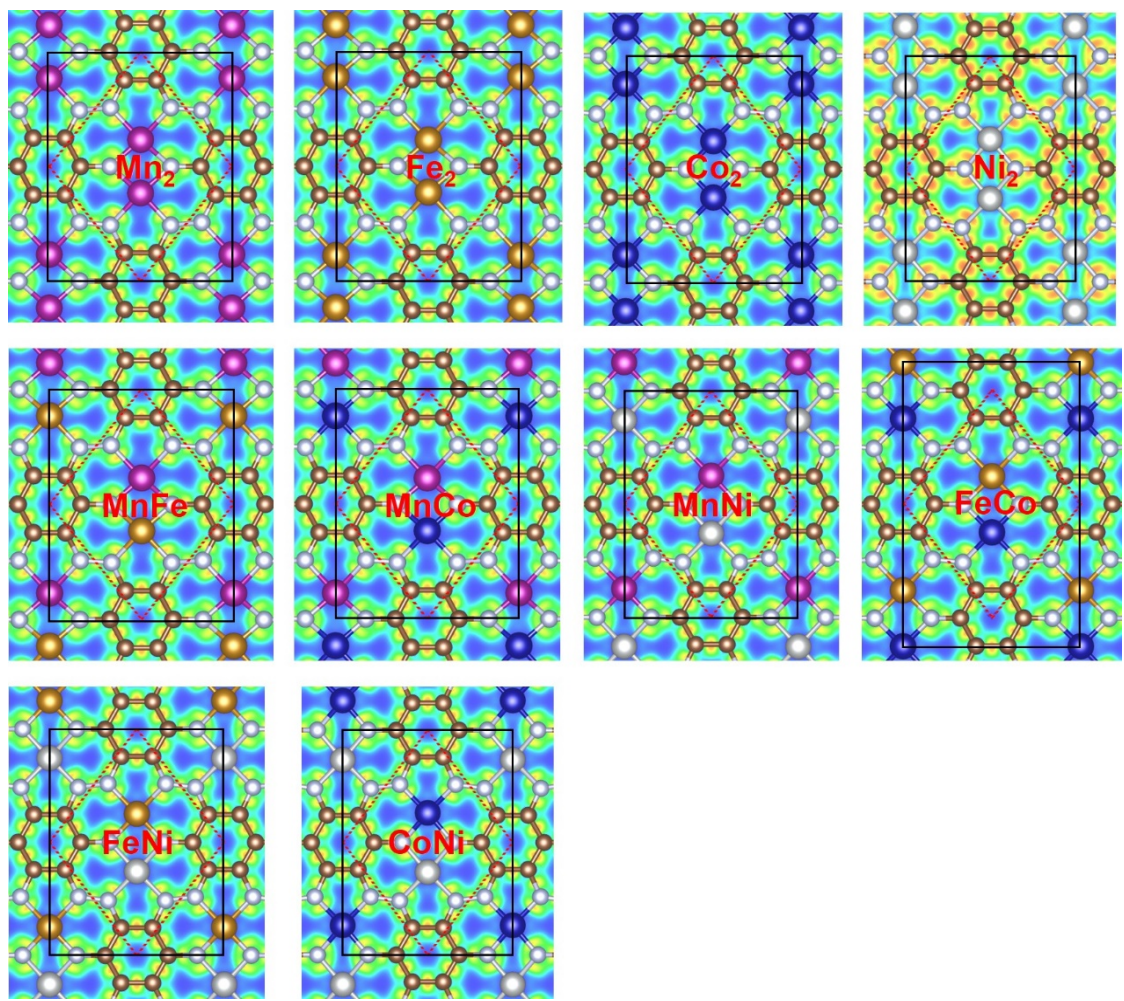

**Fig. S7.** The atomic structures with the planer-drawn electron localization function (ELF) of various UHD-DACs. The red dash lines imply the primitive cell and the black solid lines denote the conventional cell.

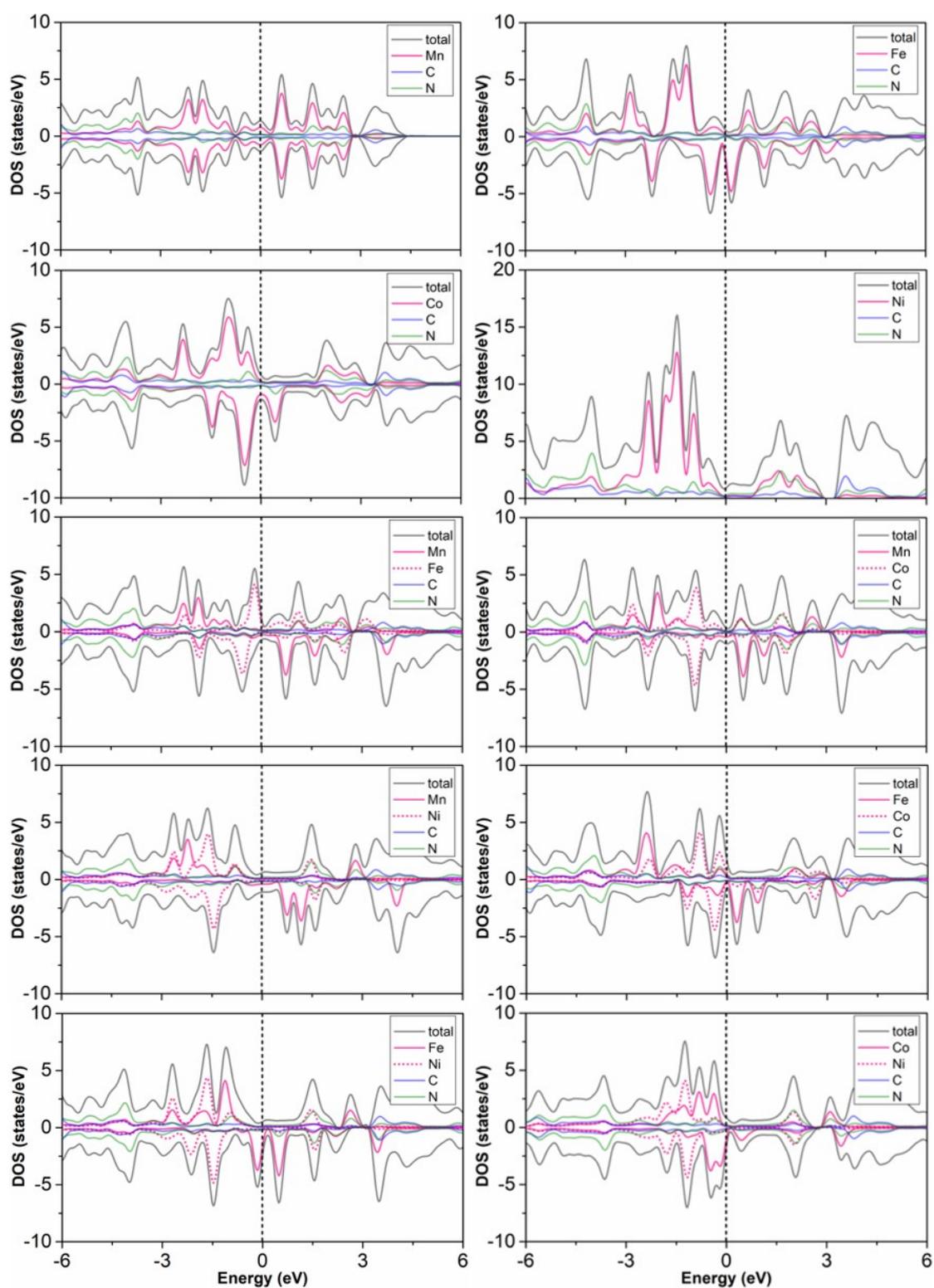

**Fig. S8.** The total density of states (TDOS), and partial density of states (PDOS) for various stable UHD-DACs based on GGA-PBE functional, respectively.

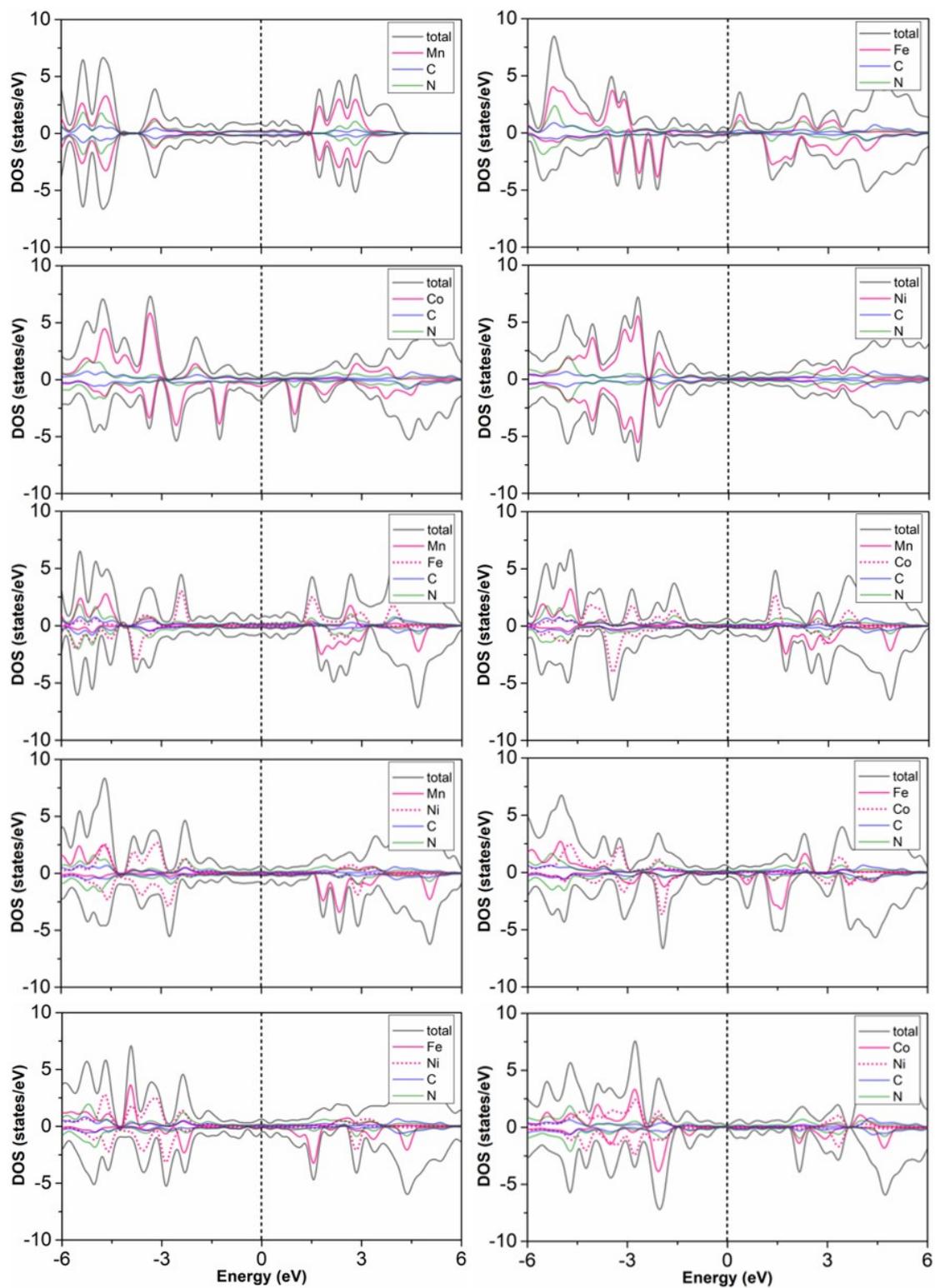

**Fig. S9.** The total density of states (TDOS), and partial density of states (PDOS) for various stable UHD-DACs based on HSE06 hybrid functional, respectively.

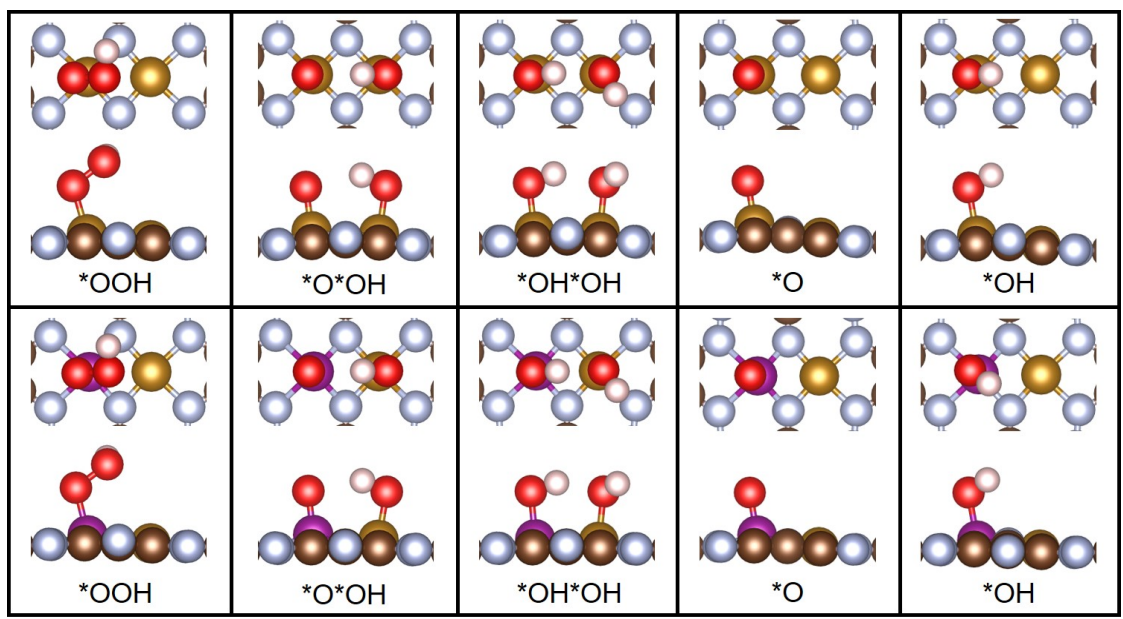

**Fig. S10.** The atomic configurations of *OOH, *O*OH, *OH*OH, *O, and *OH species on Fe$_2$N$_6$ and MnFeN$_6$ UHD-DACs, respectively. The other UHD- and LD-DACs have the same atomic configurations of oxygenated species.

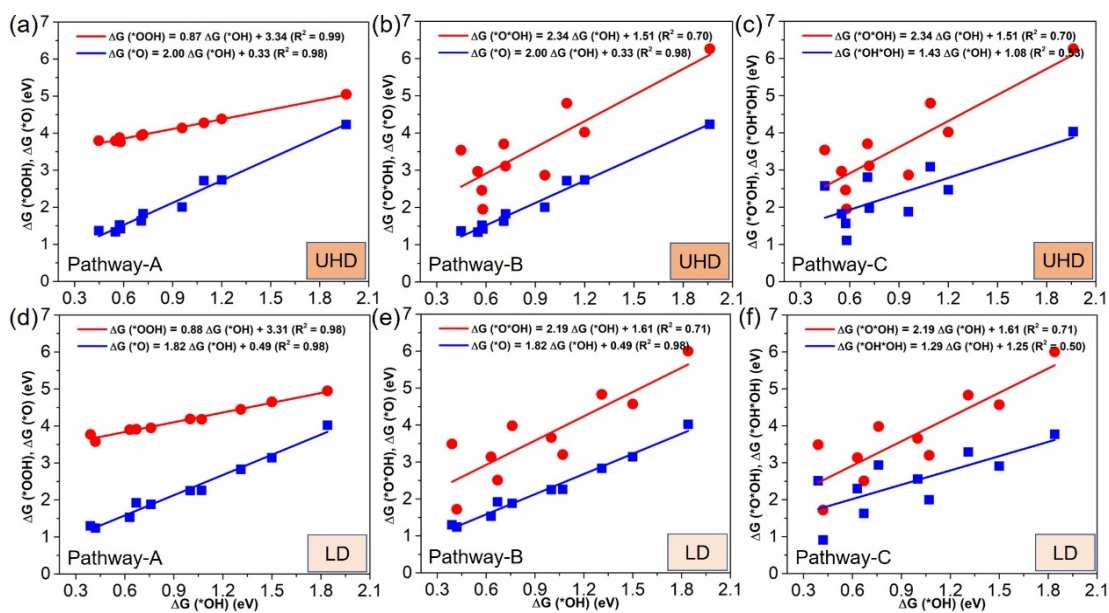

**Fig. S11.** The scaling relationship between the Gibbs adsorption free energy of the oxygenated intermediates for pathway-A, pathway-B, and pathway-C over the corresponding (a-c) UHD- and (d-f) LD-DACs, respectively. The scaling relationship for UHD-DACs are reprinted in the **Fig. 3** of main text.

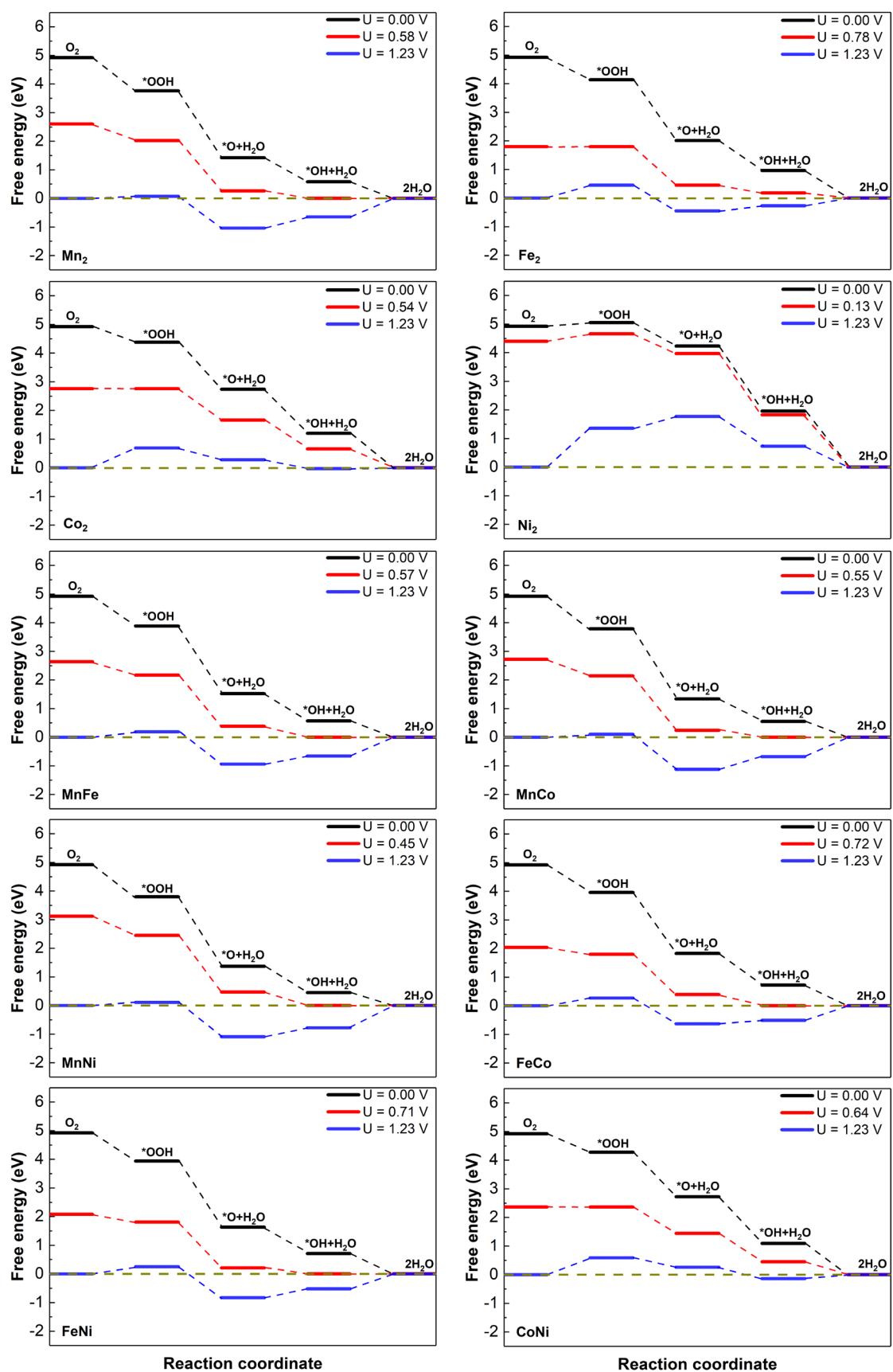

**Fig. S12.** The free energy diagrams of ORR for all UHD-DACs at different potentials for pathway-A.

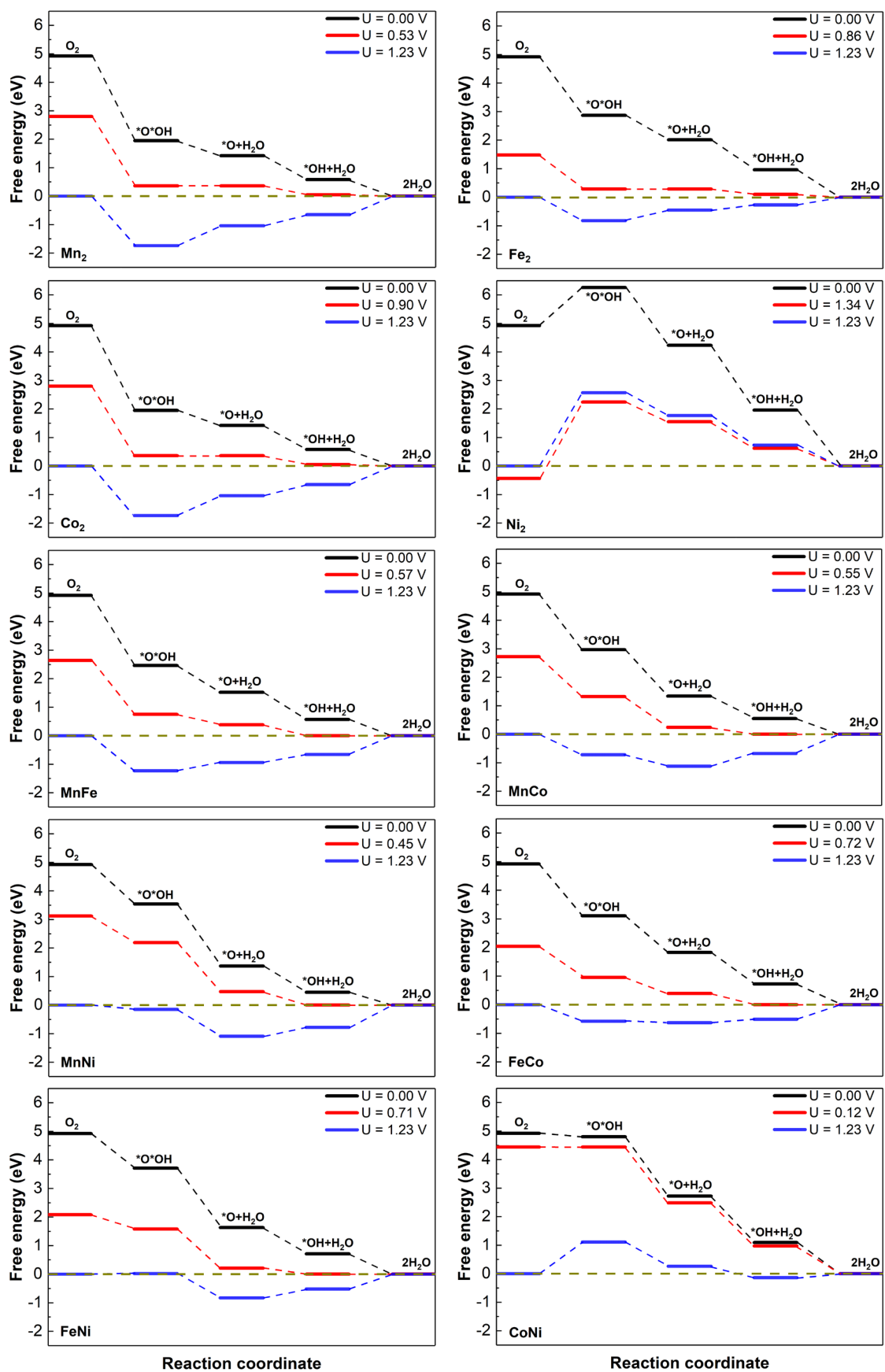

**Fig. S13.** The free energy diagrams of ORR for all UHD-DACs at different potentials for pathway-B.

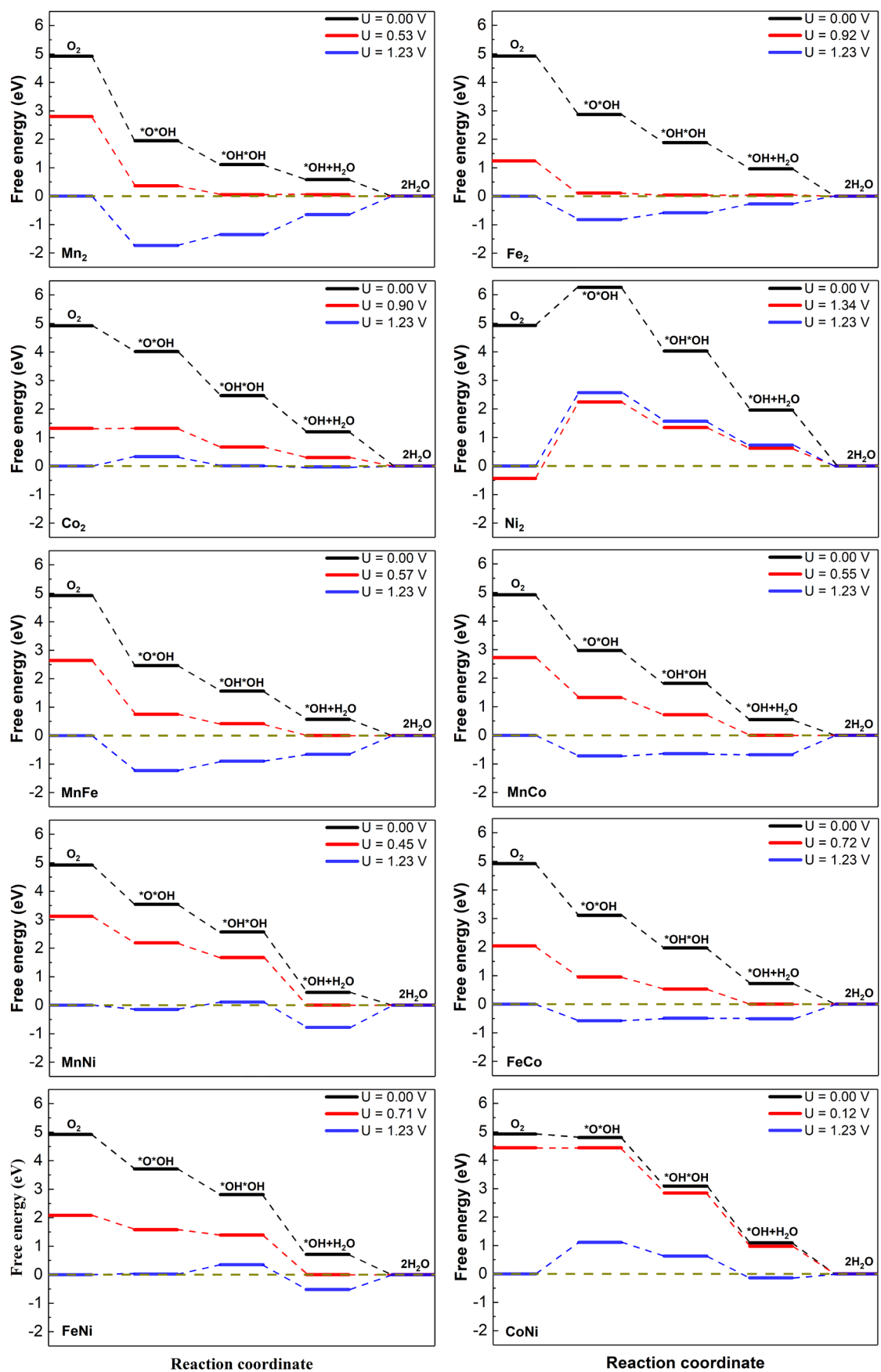

**Fig. S14.** The free energy diagrams of ORR for all UHD-DACs at different potentials for pathway-C.

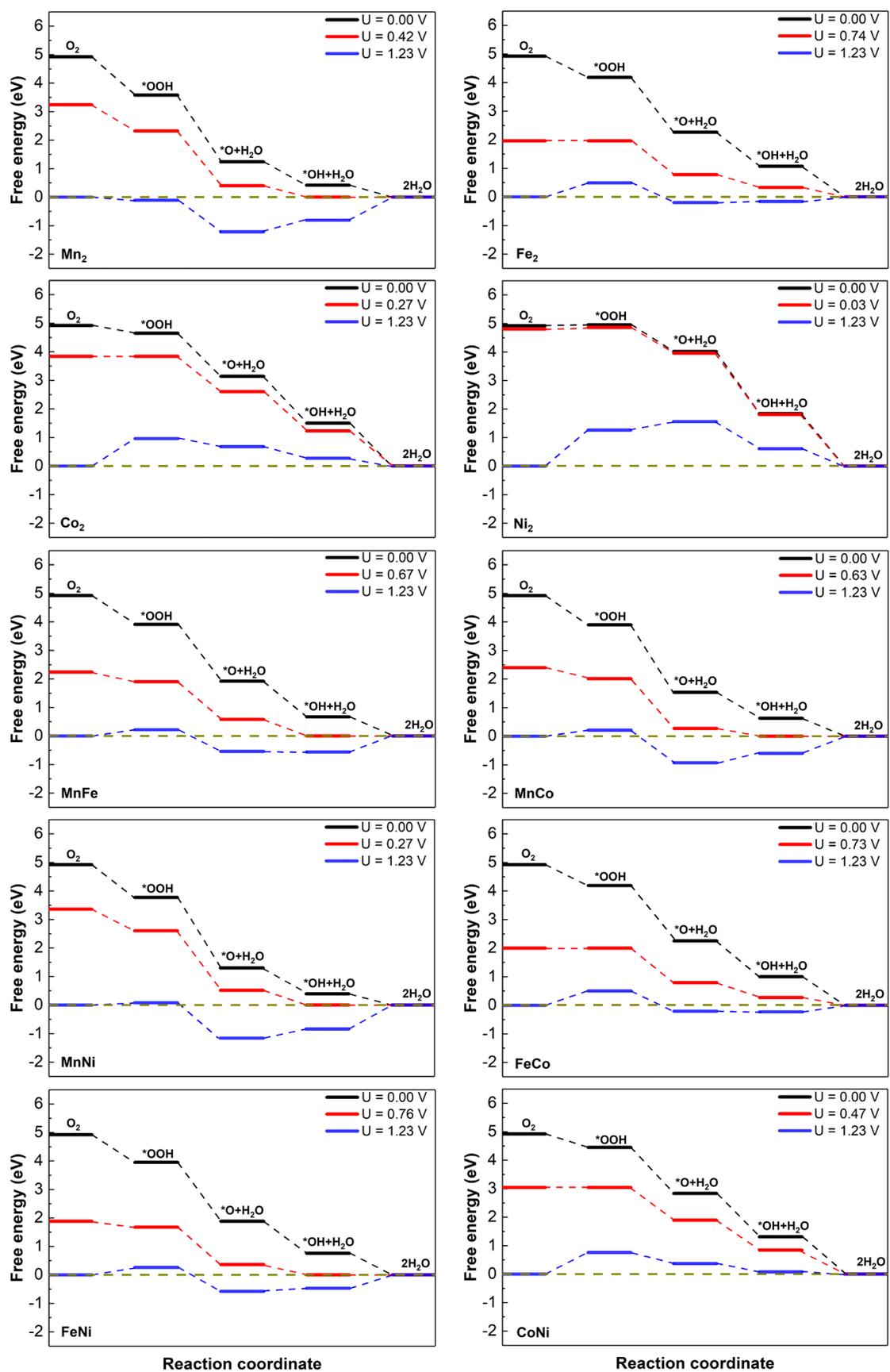

**Fig. S15.** The free energy diagrams of ORR for all LD-DACs at different potentials for pathway-A.

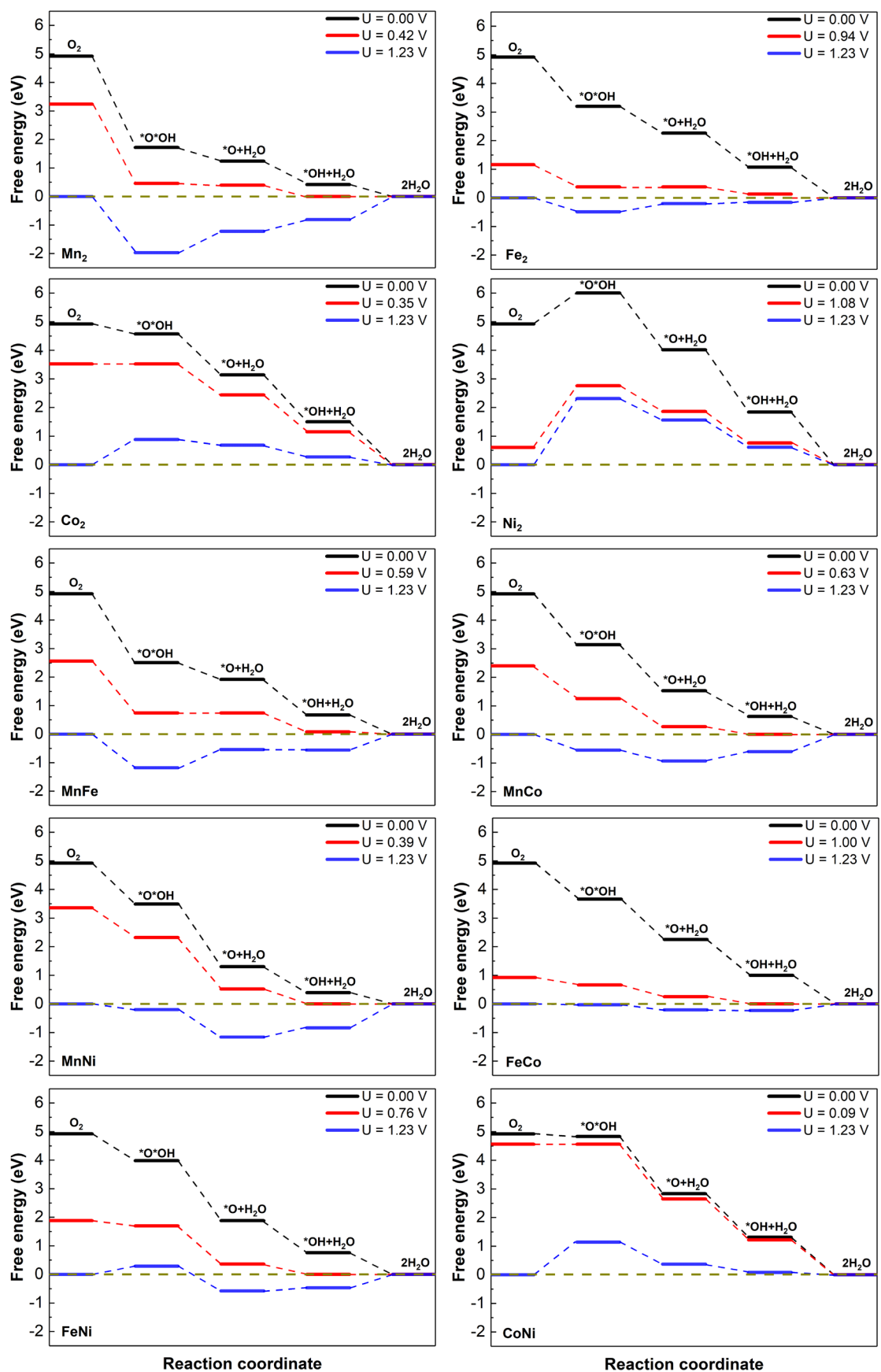

**Fig. S16.** The free energy diagrams of ORR for all LD-DACs at different potentials for pathway-B.

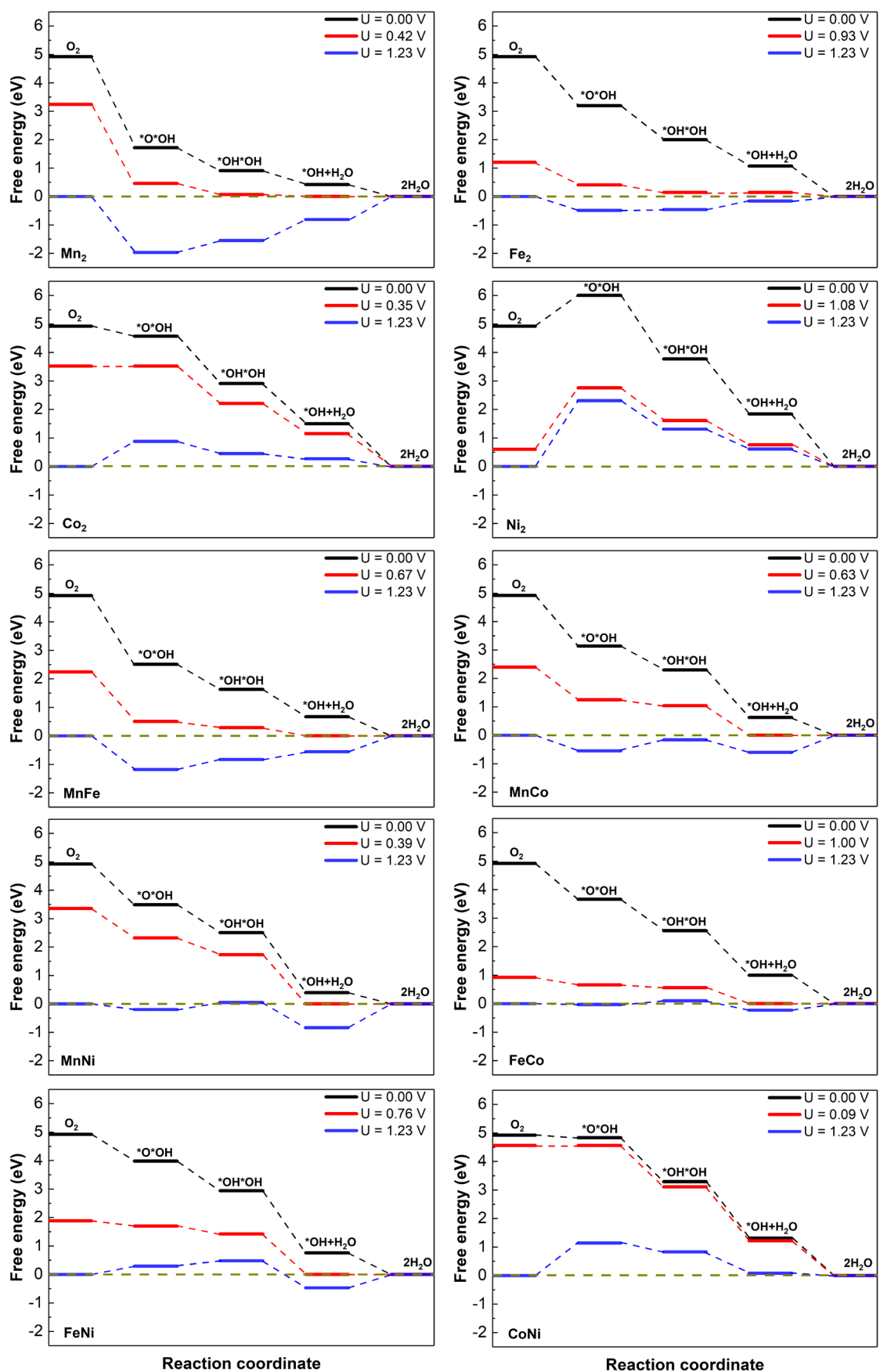

**Fig. S17.** The free energy diagrams of ORR for all LD-DACs at different potentials for pathway-C.

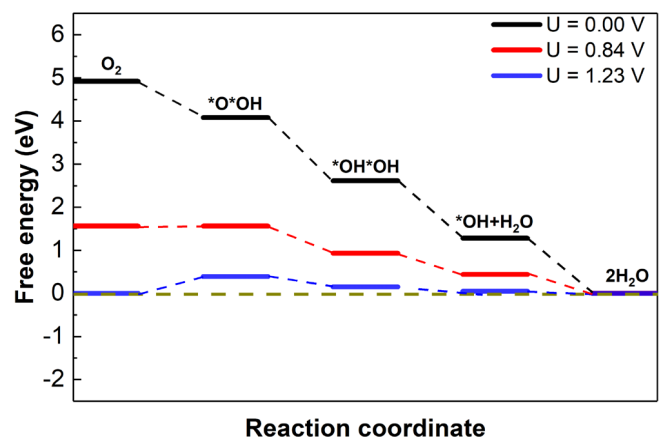

**Fig. S18.** The free energy diagrams of ORR for $Co_2N_6$ UHD-DAC at different potentials for pathway-C based on HSE06 functional.

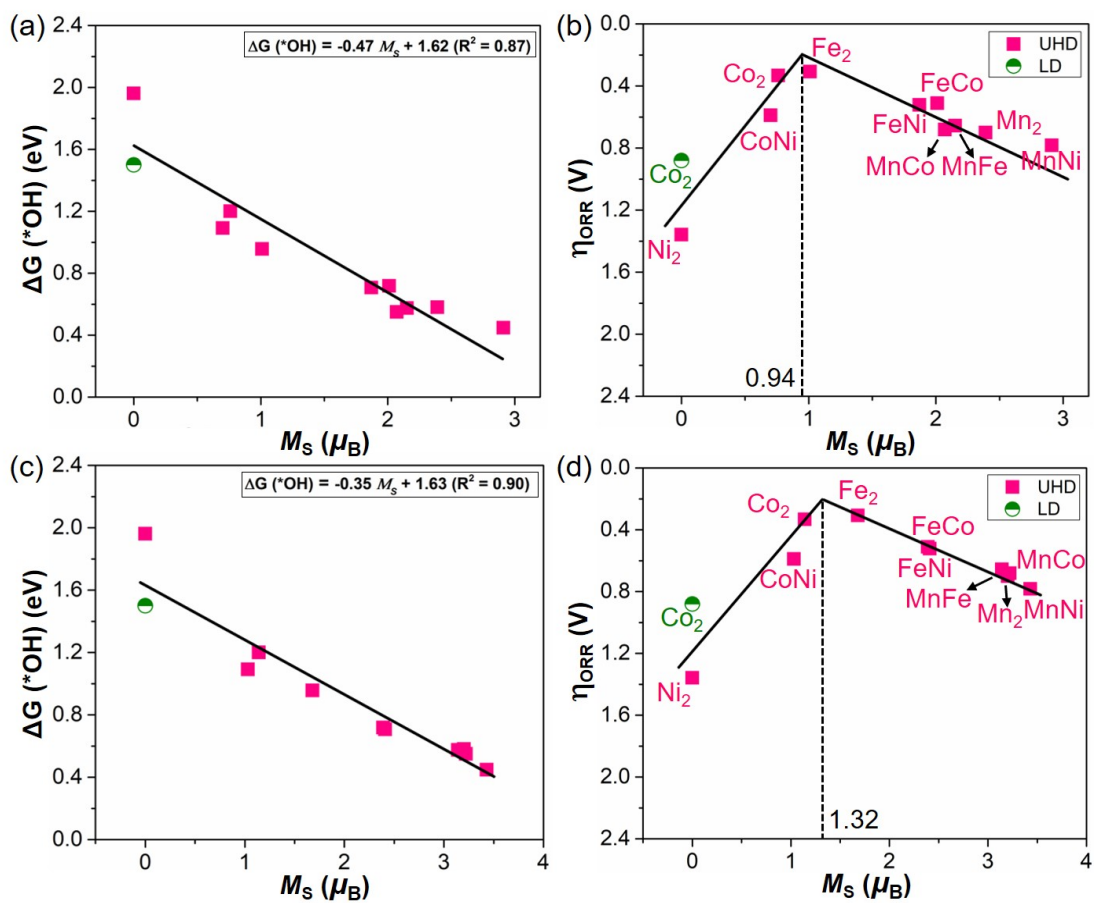

**Fig. S19**. (a)/(c) The scaling relationship between $\Delta G$ (*OH) for optimal pathway and the local spin magnetic moments ($M_S$) for the metal atom that anchor the *OH. (b)/(d) The volcano plots for the $\eta_{ORR}$ for optimal pathway as the function of $M_S$. The pink and green marks represent the UHD- and LD-DACs, respectively. The upper (a-b) and bottom (c-d) results are calculated based on PBE and HSE06 functional, respectively. In the volcano plots (b) and (d), the critical spin magnetic moment corresponding to the optimal active point is also given from PBE and HSE06 functional, respectively.

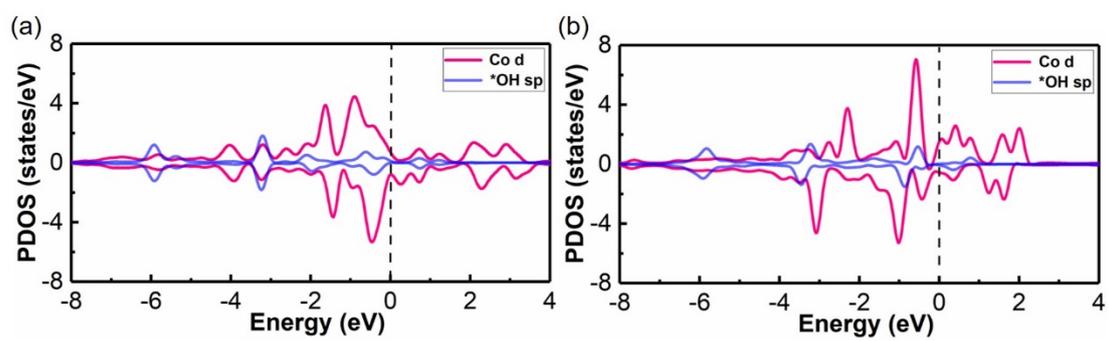

**Fig. S20.** The PDOS with electronic orbital interaction for $Co_2N_6$ (a) UHD- and (b) LD-DACs with *OH adsorption, including the Co 3*d* orbitals (pink) and *OH *sp* orbitals (blue) for DAC with OH⁻ adsorption, respectively. Here the Fermi level was set to 0.

**Note S1. Solvation effect**

Herein, the water configurations and solvation effects of the intermediates were obtained by molecular dynamics simulations (combining force-field molecular dynamic simulation implemented in LAMMPS [69] and *ab initio* molecular dynamic simulation in VASPsol [67]). The details are present below:

Firstly, we constructed a random explicit water model with a density of about 1 g/cm$^3$, including 20 H$_2$O molecules. For this explicit solvent model with 20 H$_2$O molecules, we used the LAMMPS to conduct TIP3P [70] force-field molecular dynamic (MD) simulations for 5 ns with the NVT ensemble at 300 K. Then we added this explicit water model by LAMMPS to the catalyst surface to implement the *ab initio* MD simulations (AIMD) by VASPsol for 3 ps at 300 K. The final structures are used to obtain the explicit solvation energies of reaction species. The explicit structure model with 20 explicit water molecules for various adsorbed species on Mn$_2$N$_6$ UHD-DAC are shown in **Fig. S21** and the explicit solvation energies for different adsorbed species are shown in the **Table S6**. It is found that the solvation energies from the continuum solvation model by VASPsol are comparable with those from our explicit model and those of previous work with explicit model. [71] Moreover, the explicit solvation effect doesn't alter their optimal reaction pathway for ORR on both UHD- and LD-DACs (**Table S7**). Importantly, the solvation effect has a small influence on the $\eta_{ORR}$ values, for which Fe$_2$N$_6$ and Co$_2$N$_6$ UHD-DACs still have the smallest $\eta_{ORR}$ and deliver the highest ORR activity among all the UHD-DACs. These results indicate the reasonability of our computational methods for the ORR activity.

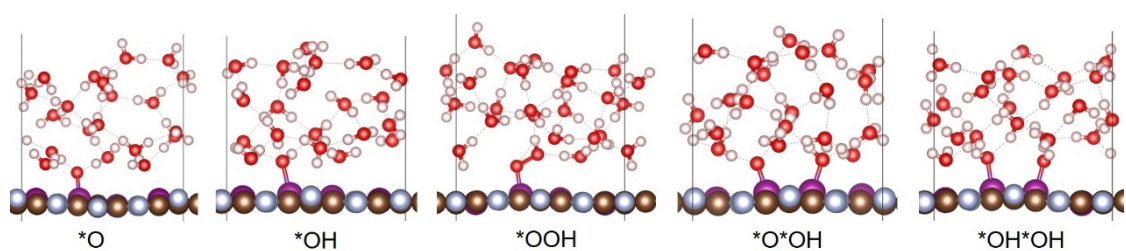

**Fig. S21**. The explicit structure model with 20 explicit water molecules for various adsorbed species on Mn$_2$N$_6$ UHD-DAC. Due to the large computational cost, here we only take Mn$_2$N$_6$ UHD-DAC as representative.

**Table S6**. The solvation energies ($\Delta E_{sol}$, eV) for different adsorbed species. The explicit structure model with 20 explicit water molecules for various adsorbed species are shown in **Fig. S21**.

| Species | $\Delta E_{sol}$ (implicit) | $\Delta E_{sol}$ (explicit) | $\Delta E_{sol}$ from Ref [18] | $\Delta E_{sol}$ from Ref [68] | $\Delta E_{sol}$ from Ref [59] | $\Delta E_{sol}$ from Ref [70] |
|---|---|---|---|---|---|---|
| *O | -0.29 | -0.27 | / | -0.29 | / | 0.36 |
| *OH | -0.22 | -0.30 | -0.19 | -0.38 | -0.30 | 0.27 |
| *OOH | -0.26 | -0.37 | / | -0.30 | -0.30 | 0.21 |
| *O*OH | -0.33 | -0.33 | -0.34 | / | / | / |
| *OH*OH | -0.31 | -0.52 | -0.40 | / | / | / |

**Table S7**. The overpotentials of ORR ($\eta_{ORR}$) with implicit and explicit solvation effect through the optimal reaction pathway (A, B or C) over corresponding UHD- and LD-DACs.

| DACs | $\eta_{ORR}$ with implicit solvation effect | | $\eta_{ORR}$ with explicit solvation effect | |
|---|---|---|---|---|
|  | UHD | LD | UHD | LD |
| Mn$_2$ | 0.70 (C) | 0.81 (C) | 0.83 (C) | 0.89 (C) |
| Fe$_2$ | 0.31 (C) | 0.29 (C) | 0.44 (C) | 0.32 (C) |
| Co$_2$ | 0.33 (C) | 0.88 (C) | 0.33 (C) | 0.88 (C) |
| Ni$_2$ | 1.36 (A) | 1.26 (A) | 1.25 (A) | 1.15 (A) |
| MnFe | 0.66 (B) | 0.56 (C) | 0.74 (B) | 0.64 (C) |
| MnCo | 0.68 (B) | 0.60 (B) | 0.76 (B) | 0.67 (B) |
| MnNi | 0.78 (B) | 0.84 (B) | 0.86 (B) | 0.91 (B) |
| FeCo | 0.51 (B) | 0.23 (B) | 0.59 (B) | 0.31 (B) |
| FeNi | 0.52 (B) | 0.47 (B) | 0.60 (B) | 0.55 (B) |
| CoNi | 0.59 (A) | 0.76 (A) | 0.48 (A) | 0.65 (A) |